\documentclass[aps,prd,showpacs,showkeys,groupedaddress,nofootinbib]{revtex4}
\usepackage{graphicx,amsmath,amssymb}
\def\be{\begin{equation}}
\def\ee{\end{equation}}
\def\bea{\begin{eqnarray}}
\def\eea{\end{eqnarray}}
\def\ba{\begin{array}}
\def\ea{\end{array}}
\def\<{\left\langle}
\def\>{\right\rangle}
\def\({\left(}
\def\){\right)}
\def\e{{\rm e}}
\def\s{$\sim$}

\begin{document}
\title{
Universality crossover between chiral random matrix ensembles\\
and twisted SU(2) lattice Dirac spectra
}
\author{Shinsuke M. Nishigaki}
\email{mochizuki@riko.shimane-u.ac.jp}
\affiliation{
Graduate School of Science and Engineering,
Shimane University,
Matsue 690-8504, Japan
}
\date{August 16, 2012; Revised: November 22, 2012}
\begin{abstract}
Motivated by 
the statistical fluctuation of Dirac spectrum of
QCD-like theories subjected to (pseudo)reality-violating perturbations
and in the $\varepsilon$ regime,
we compute the smallest eigenvalue distribution and the level spacing distribution of 
chiral and nonchiral parametric random matrix ensembles of Dyson--Mehta-Pandey type.
To this end we employ the Nystr\"{o}m-type method
to numerically evaluate the Fredholm Pfaffian of the integral kernel
for the chG(O,S)E-chGUE and G(O,S)E-GUE crossover.
We confirm the validity and universality of our results
by comparing them with several lattice models,
namely fundamental and adjoint staggered Dirac spectra of
SU(2) quenched lattice gauge theory under the
twisted boundary condition (imaginary chemical potential) or perturbed by phase noise.
Both in the zero-virtuality region and in the spectral bulk,
excellent one-parameter fitting 
is achieved already on a small $4^4$ lattice.
Anticipated scaling of the fitting parameter
with the twisting phase, mean level spacing, and the system size 
allows for precise determination of the pion decay (diffusion)
constant $F$ in the low-energy effective Lagrangian.

\end{abstract}
\pacs{
02.10.Yn, 
05.40.-a, 
11.30.Rd, 
12.38.Gc, 
12.39.Fe  
}
\keywords{Dirac eigenvalues, GOE-GUE, GSE-GUE, chiral random matrices, Fredholm determinant,
Nystr\"{o}m-type method, two-color QCD, adjoint QCD, chiral Lagrangian, pion decay constant}
\maketitle
 
\section{Introduction}
The understanding of the phase structure of fermion/gauge systems
has posed a challenge for particle and nuclear physics communities alike. 
One possible option to avoid
the sign problem that the Euclidean Dirac operator becomes
non-Hermitian and the Boltzmann weight complex in the presence of chemical potential $\mu$
is to substitute QCD by its hypothetical two-color version on a lattice \cite{nak}, 
in which the Boltzmann weight
(with pairs of degenerated quark flavors)
is positive-definite even at finite $\mu$
due to the (pseudo)reality of the representations of SU(2).
It has long been 
appreciated that the theories with quarks in the (pseudo)real representation
exhibit exotic types of
spontaneous breakdown of global flavor symmetry \cite{peskin,ls}. 
Quarks and charge-conjugated antiquarks are combined into an extended
Nambu multiplet, which in turn is expected to break down to the extended vector subgroup \cite{vw}.
Then the effect of the chemical potential that breaks the extended flavor symmetry 
is unambiguously incorporated in the low-energy effective description
through the flavor-covariant derivative \cite{kst,kogut,stv,dn}.
The assertions and/or analytic predictions, possibly based upon the effective theory,
can be quantitatively compared with the Monte Carlo simulations of SU(2) lattice gauge theory 
\cite{dagotto,baillie,hands,kogut2},
provided that the lattice regularization respects the relevant flavor symmetry group.
Accordingly, the two-color QCD has served as an insightful testing ground for the realistic chromodynamics,
as well as a tractable lattice model that is interesting by its own right.

The difference in the global symmetries is a reflection of
the difference in the antiunitary symmetries of the Dirac operators:
Dirac operators in the fundamental representation of SU(2) and in the adjoint of SU($N$) are
essentially real symmetric and quaternion self-dual, respectively,
whereas that in the fundamental of SU($N\geq3$) is merely complex Hermitian,
as Verbaarschot \cite{ver} dubbed ``threefold way" after Dyson's original proposal \cite{dys}.
Thus the inclusion of the reality- or  selfduality-violating chemical potential $\mu$ 
in two-color QCD casts itself in statistical properties of Dirac spectra.
As the symmetry-violating effect of $\mu$ in the two-color QCD
is inherent in its low-energy
effective theory and is well under control,
one can predict the fluctuation of the Dirac eigenvalues
in the $\varepsilon$ regime (i.e., below the Thouless energy)
from its zero-momentum part. This, in turn, is 
equivalent to the 
chiral Gaussian orthogonal or symplectic ensemble (chGOE, chGSE) \cite{sv, vz} 
in its non-Hermitian extended form
by the introduction of schematic (real) $\mu$ component \cite{hjv,ake}.

At this point we should note that, the reality or self-duality of the SU(2)-chromodynamic Dirac operator
could as well be violated by the inclusion of any Hermitian component in the complex representation,
most simply by U(1) electrodynamics or random phases, or even by a fixed Abelian Aharonov-Bohm (AB) flux 
background (i.e., twisted boundary condition or imaginary chemical potential) \cite{jnpz,svd,mt}.
These cases are distinct from the previously mentioned case in that
the Dirac eigenvalues stay real even after the inclusion of symmetry violations
and their statistical behavior exhibits {\em crossover}, 
rather than develops into the complex plane.
In terms of the effective $\sigma$-model description, these two cases are almost identical,
save for the difference of the sign of ${\rm tr}\,\hat{B} Q^\dagger\hat{B} Q$ term
($\mu^2F^2 $ or $(i\mu)^2 F^2$, see Sec.\ IV.).

Crossover between universality classes of Hermitian random matrix ensembles \cite{dys2,mp,meh},
namely GOE-GUE and GSE-GUE, is extensively studied 
in the context of disordered \cite{aie,asa} and quantum-chaotic Hamiltonians 
\cite{bgas,snmb,ns} with its time-reversal invariance slightly broken by 
weak magnetic field or AB flux applied \cite{dm}.
On the other hand, the chiral or superconducting variant of universality crossover
appears to be a relatively unexplored field so far.
Previous attempts in this area either focused on
the level number variance and the spectral form factor
(both of which are integral transforms of the two-level correlator 
and insensitive to chiralness)
of two-color QCD with AB fluxes
versus GOE-GUE crossover \cite{jnpz},
or have fruited in a series of tours de force by Damgaard and collaborators \cite{dam1,dam2}
devoted to the analytical computation of
the level density and individual small eigenvalue distributions
for the spectral crossover within the chiral Gaussian unitary ensemble (chGUE) class,
due to the imaginary isospin chemical potential.
Despite that analytic results for the microscopic spectral correlation functions
are known for some time for chGOE-chGUE and chGSE-chGUE crossover \cite{nag,kt},\footnote{
Both of Refs.~\cite{nag,kt}, which practically computed $n$-level correlators of parametric chiral random matrices
in the Pfaffian form,
cited Ref.~\cite{vz} by mentioning that 
``chiral random matrices serve as effective models of lattice gauge theory,
namely QCD" [translation from the former] 
and ``three chiral versions of random matrix ensembles in the particle physics of QCD" [quote from the latter].
Thus it would be fair to presume that these authors have envisaged possible application of their results
toward the crossover phenomena in the QCD Dirac spectrum.}
they are yet to encounter with physical application.
To the best of our knowledge, the only example of the crossover involving 
{\em different} Hermitian chiral universality classes discussed in a physical setting is
the CI-C transition for the 
normal/superconducting hybrid interface in a magnetic field \cite{koz}.
Part of the aim of this paper is to present novel examples of the
physical application of the crossover between chiral Hermitian universality classes
in the realm of lattice gauge theory, namely of QCD-like theories.

In this paper we shall show that such Dirac spectra
indeed exhibit symmetry crossover from the 
chGOE or chGSE to the chGUE universality class,
precisely as predicted by the chiral variants of
parametric random matrix ensembles of Dyson \cite{dys2} and Mehta-Pandey \cite{mp,meh}, 
both in the zero-virtuality region and in the spectral bulk.
Through the excellent one-parameter fit to the parametric random matrix results,
we extract the pion decay (diffusion) constant
in the effective Lagrangian.
As our method adopts 
the level spacing and the smallest eigenvalue distributions
that are extremely sensitive to the fitting parameter as primary fitting observables
(see Refs.\cite{lbhw,bbms,ai} for recent efforts along this line),
it enjoys a clear advantage over 
the methods using $n$-level correlation functions,
and presents promising applications in analyzing the numerical data of QCD-like theories.

This paper is composed as follows: In Sec.\ II 
we briefly review the universality crossover in the spectrum of random matrices.
Then we shall 
compute and plot the level spacing distribution in the bulk 
and the smallest eigenvalue distribution at the hard edge (origin) of the spectrum using
the Nystr\"{o}m-type method, the latter 
being our new contribution.
In Sec.\ III we measure the level spacing and smallest eigenvalue distributions
of fundamental and adjoint staggered Dirac operators
of SU(2) 
quenched lattice gauge theory
with weak AB flux or phase noise included,
and fit the spectral data with
the predictions of (ch)GSE-(ch)GUE and (ch)GOE-(ch)GUE
crossover. 
We show that an excellent one-parameter fitting can be achieved
for all of our cases of concern, so that 
an accurate determination of the crossover parameter is possible.
In Sec.\ IV
we extract the pion decay constant in the effective chiral Lagrangian
from the flux dependence of the crossover parameter.
We shall conclude in Sec.\ V with some discussions on the two-color QCD+QED
simulation and on possible directions of future study.

\section{Parametric (chiral) random matrices}
Transition of spectral fluctuation
from one universality class to another with a different antiunitary symmetry
has been proven to occur both in quantum-chaotic and disordered Hamiltonians
with weakly broken time-reversal invariance.
The result is universal in a sense that the local spectral fluctuation is 
sensitive only to a single crossover parameter $\rho$ defined below.
The reason for this universality is 
traced back to
the nonlinear $\sigma$ model governing the spectral statistics,
which can either be derived
by the conventional disorder averaging of random Hamiltonians \cite{aie} or
by the summation over Sieber-Richter--like encountering multiplet of periodic orbits
of chaotic dynamical systems \cite{snmb},
completely irrespective of the details of dynamics.
Accordingly, one can resort to the simplest model which yields the identical $\sigma$ model, i.e.\ 
parametric random matrix ensembles. 
Below, we collect established results 
on the spectral correlation of the parametric random matrices of nonchiral and chiral types for completeness and
refer the reader to the original references 
\cite{dys2,mp,meh,nag}
for their derivations.

\subsection{GOE-GUE and GSE-GUE crossover}
We consider an ensemble of $N\times N$ Hermitian complex (quaternion) matrices $H=H_{\rm S}+i\alpha H_{\rm A}$,
with $H_{\rm S}$ real symmetric (quaternion self-dual) 
and $H_{\rm A}$ real antisymmetric (quaternion anti-self-dual) matrices
distributed according to Gaussian measures of variance $\sigma^2$.
This parametric (also called Brownian-motion or dynamical) random matrix 
ensemble interpolates between the two limiting cases,
GOE (GSE) at $\alpha=0$ and GUE at $\alpha=1$.
Take a point $\lambda$ from the bulk part of the spectrum of $H$ and denote
the mean level spacing around $\lambda$ by $\varDelta(\lambda)$.
Then the $n$-point correlation function of the eigenvalues $\{\lambda_i\}$ of $H$
in the vicinity of $\lambda$ is given,
in the limit $N\to\infty,\ \alpha\to 0$ and 
$\rho\equiv {\alpha \sigma}/{\varDelta(\lambda)}$ fixed
 (we follow Mehta's book \cite{meh} for the definition of $\rho$),
as a Pfaffian,
\bea
&&R_{n}(x_1,\ldots,x_n)={\rm Pf}\bigl(Z \left[K(x_i, x_j)\right]_{i,j=1}^n\bigr)=
\sqrt{\det \left[K(x_i, x_j)\right]_{i,j=1}^n}~,
\nonumber\\
&&
K(x, y)=\left[
\ba{cc}
S(x,y) & I(x, y)\\
D(x,y) & S(y, x)
\ea
\right],\ \ 
Z=i\sigma_2 \otimes \openone_n,
\label{Pfaffian}
\eea
where $x_i\equiv\lambda_i/\varDelta(\lambda)$ are the unfolded eigenvalues, and 
$S(x,y), D(x,y), I(x,y)$ as functions of $r=x-y$ are given by
\bea
\mbox{[GOE-GUE]}&&
S(r)=\frac{\sin \pi r}{\pi r}~,~~
D(r)=\frac1\pi \int_0^\pi dv\,v\,\e^{2\rho^2 v^2}\sin v r~,~~
I(r)=\frac1\pi\int_\pi^\infty \frac{dv}{v}\e^{-2\rho^2 v^2}\sin  v r~,
\label{GOEGUE}\\
\mbox{[GSE-GUE]}&&
S(r)=\frac{\sin \pi r}{\pi r}~,~~
D(r)=\frac1\pi \int_\pi^\infty dv\,v\,\e^{-2\rho^2 v^2}\sin v r~,~~
I(r)=\frac1\pi\int_0^\pi \frac{dv}{v}\e^{2\rho^2 v^2}\sin  v r~.
\label{GSEGUE}
\eea
The probability $E(s)$ that an interval of width $s$ contains no eigenvalue 
is then given as the Fredholm Pfaffian or square root of the Fredholm determinant \cite{mp}:
\be
E(s)=\sum_{n=0}^\infty \frac{(-1)^n}{n!} \int_0^s dx_1\cdots\int_0^s dx_n\, R_n(x_1,\ldots x_n)=
\sqrt{{\rm Det} (\openone-\hat{K}_s)}
\label{FredholmPfaffian}
\ee
where 
$\hat{K}_s$ 
is an integral operator of convoluting with the ``dynamical" sine kernel $K(x,y)$
(\ref{Pfaffian}), (\ref{GOEGUE}), (\ref{GSEGUE}), 
restricted to the interval $[0,s]$.
The probability distribution $P(s)$ of level spacings $s=x_{i+1}-x_i$
is given by its second derivative $P(s)=E''(s)$.

We should emphasize that for parametric random matrix ensembles,
local correlations of {\it unfolded} eigenvalues in the vicinity of $\lambda$
still depend upon the mean level spacing $\varDelta(\lambda)$ of the eigenvalue window in concern
through the parameter $\rho$.
Accordingly, if the parameter $\alpha$ is adiabatically increased from zero to unity,
universality crossover from GOE or GSE to GUE
takes place at a different rate in each window in the spectrum
(the denser the eigenvalues, the faster the speed of crossover).
Also note that the universal intermediate behavior of spectral fluctuations
appears only in a double limit, where the system size tends to infinity and
the $\alpha$ parameter to zero in a correlated manner; a simple thermodynamic limit $N\to\infty$
with $\alpha$ fixed
would drive the whole spectrum to the GUE class.

\subsection{chGOE-chGUE and chGSE-chGUE crossover}
The chiral version of the parametric random matrix ensembles is simply obtained by
setting the ${N}/{2}\times {N}/{2}$ block-diagonal parts of 
$H=H_{\rm S}+i\alpha H_{\rm A}$ to zero.
Accordingly, the matrix in concern takes the form
\be
H=
\(
\ba{cc}
0&H_1+i\alpha H_2\\
(H_1-i\alpha H_2)^{T,D} &0
\ea
\),\ \ \ H_1, H_2 : \frac{N}{2}\times \frac{N}{2}\ \mbox{(quaternion-)real matrices},
\ee
distributed according to Gaussian measures of variance $\sigma^2$.
This ensemble interpolates between the two limiting cases,
chGOE (chGSE) at $\alpha=0$ and chGUE at $\alpha=1$.
Since the nonzero eigenvalues of $H$ occur in the $\pm$ pairs of equal magnitude,
if suffices to retain only non-negative eigenvalues.
The $n$-point correlation function of the
eigenvalues $\{\lambda_i\}$ of $H$
in the vicinity of the origin 
is similarly expressed,
in the limit $N\to\infty,\ \alpha\to 0$ and 
$\rho\equiv {\alpha \sigma}/{\varDelta(0)}$ fixed,
as a Pfaffian (\ref{Pfaffian}) with $S, D, I$ given by
(after substituting $X\to \pi^2 x^2, Y\to \pi^2 y^2$ into the ``Laguerre-type'' formulas  
(7.2.28$\sim$56) of Ref.~\cite{nag} and multiplying them by $2\pi\sqrt{xy}$),
\bea
\mbox{[chGOE-chGUE]}
&&S(x,y)=\pi  \sqrt{x y} \left\{
\frac{x J_1(\pi  x) J_0(\pi  y)-J_0(\pi  x)y J_1(\pi  y)}{x^2-y^2}
+\frac{J_0(\pi y)}{2}\int_\pi^\infty dv\,\e^{-\rho^2 (v^2-\pi^2)}J_0(v x)\right\}
~,\nonumber\\
&&D(x,y)=
-\frac{\sqrt{x y} }{2}  \int_0^\pi dv\, v^2 \,\e^{2 \rho^2 v^2} 
\left\{x J_1(v x) J_0(v y)- J_0(v x)y J_1(v y)\right\}     
~,\nonumber\\
&&I(x,y)=
\frac{\sqrt{xy}}{2}  \int_\pi^\infty dv\,v  \int_1^\infty du\,\e^{-\rho^2 v^2(1+u^2)}
\left\{J_0(v u x) J_0(v y)- J_0(v x) J_0(v u y)\right\}
~,\label{chGOEchGUE}\\
\mbox{[chGSE-chGUE]}
&&S(x,y)=\pi  \sqrt{x y} \left\{
\frac{x J_1(\pi  x) J_0(\pi  y)-J_0(\pi  x)y J_1(\pi  y)}{x^2-y^2}
-\frac{J_0(\pi x)}{2}\int_0^\pi dv\,\e^{\rho^2 (v^2-\pi^2)}J_0(v y)\right\}
~,\nonumber\\
&&D(x,y)=
\frac{\sqrt{xy} }{2} \int_0^\pi dv\,v  \int_0^1 du\,\e^{\rho^2 v^2(1+u^2)}
\left\{J_0(v u x) J_0(v y)- J_0(v x) J_0(v u y)\right\}
~,\nonumber\\
&&I(x,y)=
\frac{ \sqrt{x y}}{2}  \int_\pi^\infty dv\, v^2 \,\e^{-2 \rho^2 v^2} 
\left\{x J_1(v x) J_0(v y)- J_0(v x)y J_1(v y)\right\}
~,\label{chGSEchGUE}
\eea
where $x_i\equiv\lambda_i/\varDelta(0)$ are the unfolded eigenvalues  (see also \cite{kt}).

The probability $E(s)$ that no eigenvalue is smaller than $s$
is again given as the Fredholm determinant (\ref{FredholmPfaffian}),
where $\hat{K}_s$ in this case is 
an integral operator of convoluting with the dynamical Bessel kernel $K(x,y)$ 
(\ref{Pfaffian}), (\ref{chGOEchGUE}), (\ref{chGSEchGUE}),
restricted to the interval $[0,s]$.
The probability distribution $p_1(s)$ of the unfolded smallest eigenvalue $s=\lambda_1/\varDelta(0)$,
which is one half of the very central level spacing, 
is then given by its first derivative $p_1(s)=-E'(s)$.

\subsection{Nystr\"{o}m-type method}
A simple but exceptionally efficient way of evaluating
the Fredholm determinant
of a trace-class integral operator $\hat{K}_s$ acting on 
the Hilbert space of $L^2$ functions $f$ over
an interval $[0,s]$ by
$(\hat{K}_s f)(x)=\int_0^s dy\,K(x,y)f(y)$,
such as the one with the sine kernel $S(r)$, 
is the Nystr\"{o}m-type method \cite{nys,bor}.
It simply discretizes the Fredholm determinant:
\be
{\rm Det}(1-\hat{K}_s)\simeq \det \left[\delta_{ij}-K(x_i,x_j) \sqrt{w_i\,w_j}\right]_{i,j=1}^m~.
\label{Nystrom}
\ee
Here the quadrature rule $\{x_1, \ldots, x_m; w_1, \ldots, w_m\}$ consists of
 a set of points $\{x_i\}$ 
taken from the interval $[0,s]$ and of positive weights $\{w_i\}$ such that
${\displaystyle \int_0^s f(x)dx  \simeq \sum_{i=1}^m f(x_i) w_i}$.
Efficient choices for the quadrature rule are
Gauss (sampling at the Legendre nodes) and Clenshaw-Curtis (sampling at the Chebyshev nodes),
for which the computational cost grows optimally as $O(m^2)$ and $O(m \log m)$, respectively \cite{tre}.
As the order $m$ of the approximation increases, 
the rhs of (\ref{Nystrom}) is proven to uniformly converge to its lhs.
The convergence is rapid and exponentially fast; the approximation error decays as $O(\e^{-{\rm const.}m})$
\cite{bor}.
In this paper we choose the Gauss quadrature rule, as 
15-digit accuracy is already attainable only with $m=5$
for the Fredholm determinant $E(0.1)$ for the sine kernel.
An extension to the matrix-valued kernel (\ref{Pfaffian}) is trivial: one merely
takes the determinant over the matrix indices as well.
Practical significance of the method in the context of random matrices and stochastic processes
is recently reappreciated and stressed in Ref.~\cite{bor}.

We have applied this Nystr\"{o}m-type method  to the dynamical sine kernel 
(\ref{Pfaffian}), (\ref{GOEGUE}), (\ref{GSEGUE}) 
and the dynamical Bessel kernel (\ref{Pfaffian}), (\ref{chGOEchGUE}), (\ref{chGSEchGUE}) 
to obtain $P(s)$ and $p_1(s)$ for G(O,S)E-GUE and chG(O,S)E-chGUE crossover, respectively.
In order to achieve accuracy that is needed for computing the first or second derivatives
($p_1(s)$ or $P(s)$) to a good precision, we have chosen the approximation order $m$ to be (at least)
20 for the former and 100 for the latter
and confirmed the stability of the results under the increment of $m$.
Numerical results 
for the region $0\leq s\leq 3\sim4$ and for the parameter range $\rho\lesssim 1$
are exhibited in Figs.~1 and 2 (left) [$P(s)$ for nonchiral random matrices]
and in Figs.~3 and 4 (left) [$p_1(s)$ for chiral random matrices]\footnote{
Due to Kramers degeneracy, a half of all the level spacings of GSE are zero
and its distribution has a peak at the origin, $P_{\rm GSE}(s\approx 0)\sim\delta(s)/2$. 
Thus, its smooth part is rescaled and normalized as $\int_{+0}^\infty P_{\rm GSE}(s)ds=1/2$.
Accordingly, $P(s)$ for GSE-GUE crossover becomes peaky near $s\approx 0$ as
the $\rho$ parameter is decreased. 
It also leads to the loss of accuracy of the Nystr\"{o}m approximation at very small $\rho$.}
Although Mehta-Pandey (in their second paper of Ref.~\cite{mp}) 
have expressed the Fredholm determinant for the dynamical sine kernel
in terms of eigenvalues of an infinite-dimensional matrix (each matrix element of which is
an integral involving prolate spheroidal functions),
these numerical plots of $P(s)$ and $p_1(s)$ for 
(namely the chiral version of) parametric random matrix ensembles do not seem to have appeared explicitly
in the literature, to the best of our knowledge%
\footnote{The first reference of \cite{dm} did not use the analytic form of $P^{(\rho)}(s)$
for GOE-GUE crossover plotted in Fig.1 (left), but employed the Wigner-surmised form (see also \cite{abps}).}.
Also plotted in the figures are 
$P(s)$'s for the two limiting cases $\rho=0$ and $\rho=\infty$, 
i.e.~for GOE (GSE) and GUE obtained by Jimbo-Miwa-M\^{o}ri-Sato \cite{JMMS}
in terms of a solution to the Painlev\'{e} V
transcendental equation subjected to an appropriate boundary condition,
and $p_1(s)$'s for chGOE (chGSE) and chGUE \cite{damn},
\be
p_1(s)
=-\frac{d}{ds}
\left\{
\ba{cll}
\exp\(-\frac{\pi}{2} s\) \!&\! \exp\(-\frac{\pi^2}{8} s^2\)
~& \mbox{[chGOE]}\\
\cosh \frac{\pi}{2}s \!&\! \exp\(-\frac{\pi^2}{8} s^2\)
~& \mbox{[chGSE]}\\
\!&\!\exp\(-\frac{\pi^2}{4} s^2 \)
~& \mbox{[chGUE]}
\ea
\right.
~.
\label{p1s}
\ee

\begin{figure}
\includegraphics[bb=0 0 252 164]{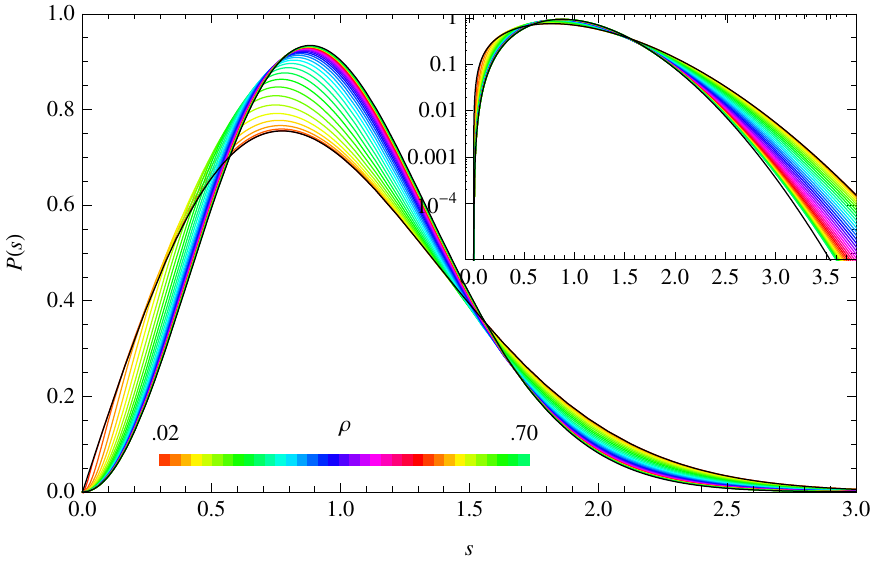}~~\includegraphics[bb=0 0 252 162]{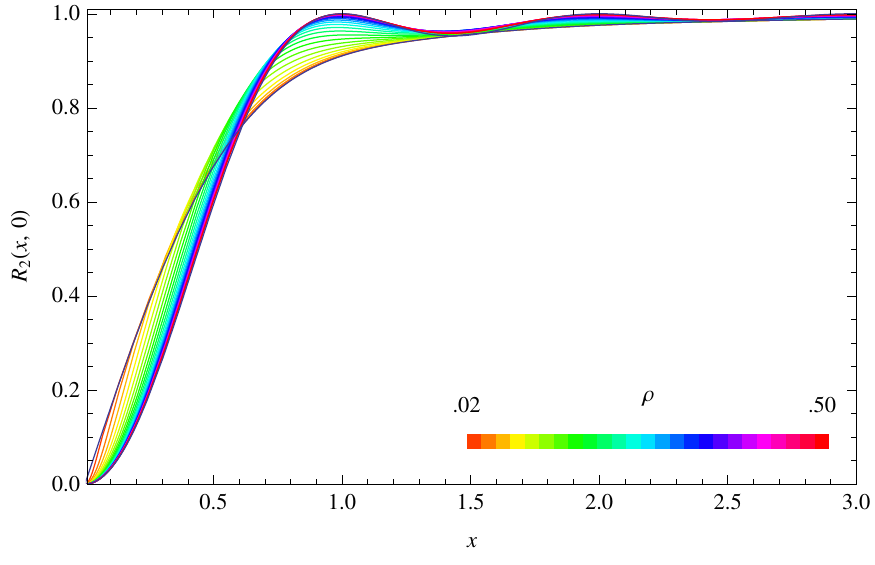}
\caption{Level spacing distribution $P(s)$ and two-level correlation function $R_2(x,0)$ for GOE-GUE crossover.
The $\rho$ parameter ranges $0.02\leq \rho \leq 0.70$ or $0.50$ by step 0.02.
The two bounding curves correspond to the GOE and GUE limits, $\rho=0$ and $\infty$.
}
\vspace{4mm}
\includegraphics[bb=0 0 252 166]{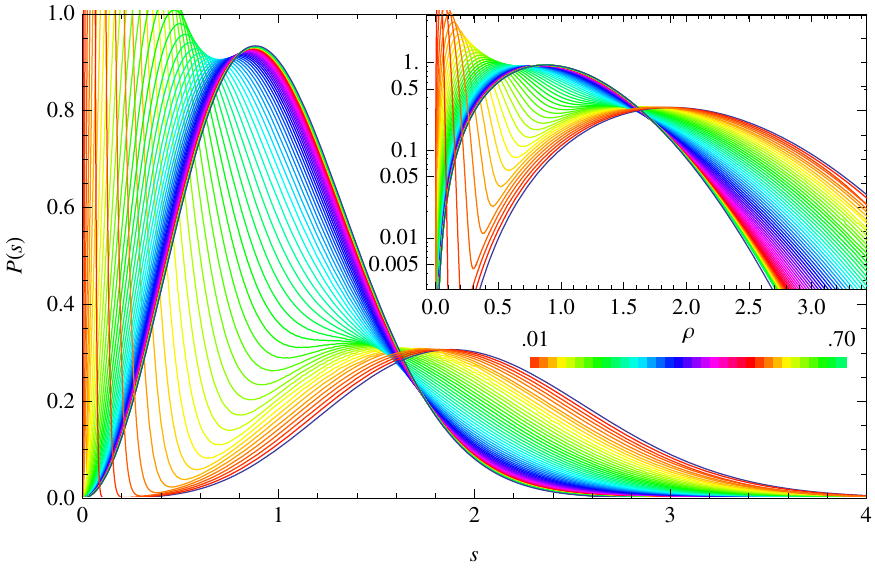}~~\includegraphics[bb=0 0 252 161]{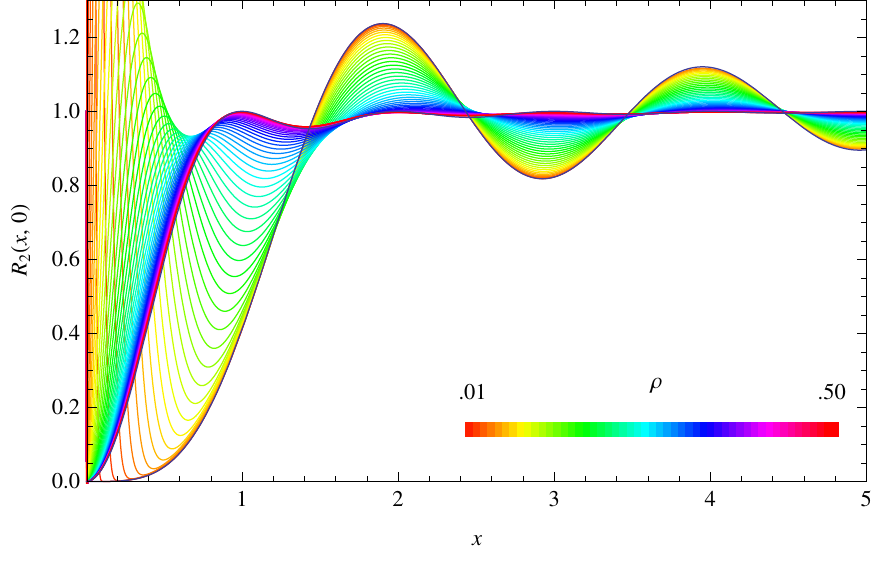}
\caption{Level spacing distribution $P(s)$ and two-level correlation function $R_2(x,0)$ for GSE-GUE crossover.
The $\rho$ parameter ranges and $0.01\leq \rho \leq 0.70$ or $0.50$ by step 0.01.
The two bounding curves correspond to the GSE and GUE limits, $\rho=0$ and $\infty$.
}
\vspace{4mm}
\includegraphics[bb=0 0 252 159]{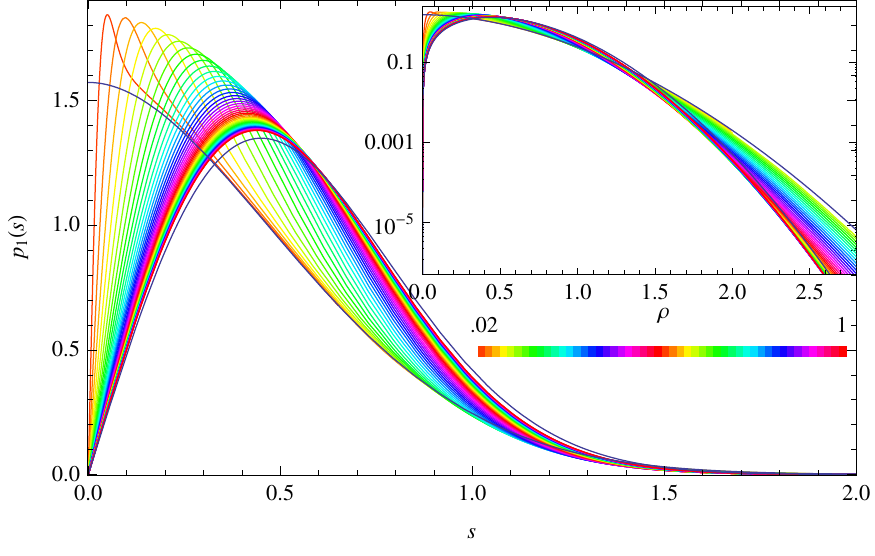}~~\includegraphics[bb=0 0 252 159]{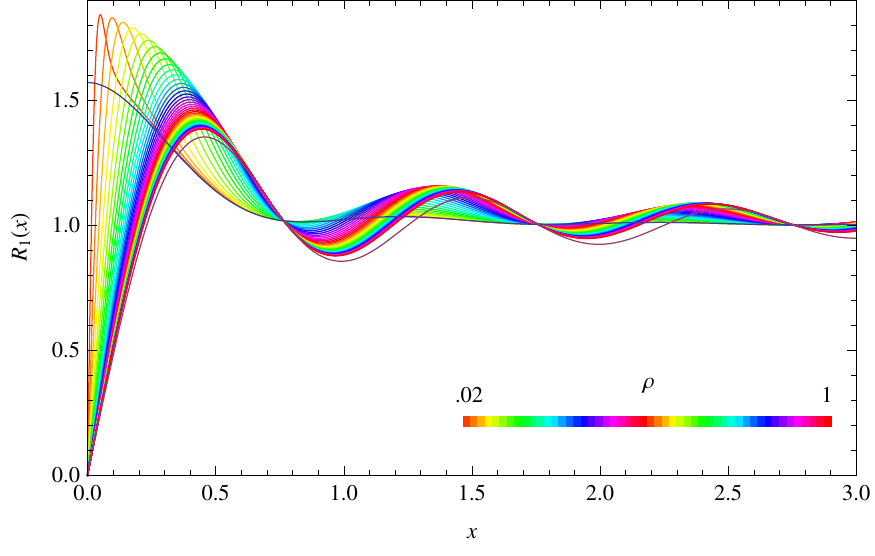}
\caption{Smallest eigenvalue distribution $p_1(s)$ and microscopic level density $R_1(x)$ for chGOE-chGUE crossover.
The $\rho$ parameter ranges $0.02\leq \rho \leq 1.00$ by step 0.02.
The two bounding curves correspond to the chGOE and chGUE limits, $\rho=0$ and $\infty$.
}
\end{figure}

\begin{figure}
\includegraphics[bb=0 0 252 159]{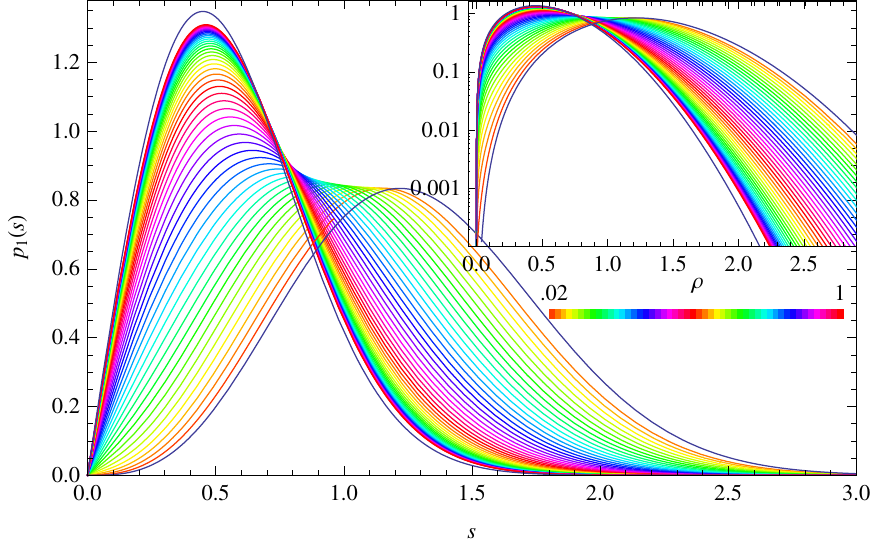}~~\includegraphics[bb=0 0 252 161]{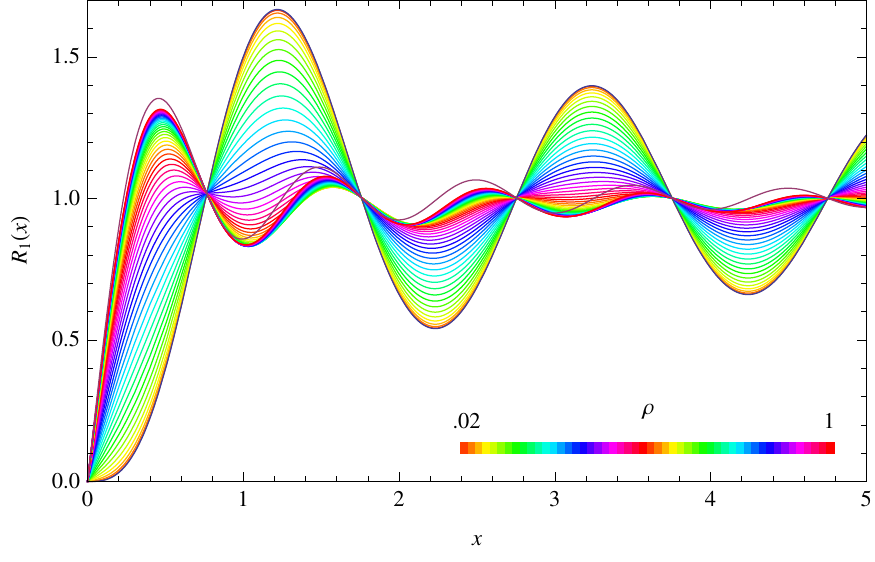}
\caption{Smallest eigenvalue distribution $p_1(s)$ and microscopic level density $R_1(x)$ for chGSE-chGUE crossover.
The $\rho$ parameter ranges $0.02\leq \rho \leq 1.00$ by step 0.02.
The two bounding curves correspond to the chGSE and chGUE limits, $\rho=0$ and $\infty$.
}
\end{figure}

We immediately observe the asymptotic behaviors of $P(s)$ and $p_1(s)$
for parametric (chiral) random matrices at finite $\rho$, 
\be
P^{(\rho)}(s)\sim C_\rho s^2,\ \ p^{(\rho)}_1(s)\sim c_\rho s^1\ \ (s\ll1);\ \ \ \ 
\log P^{(\rho)}(s),\ \log p_1^{(\rho)}(s) \sim -\gamma_\rho s^2\ \ (s\gg1).
\label{asympt}
\ee
Here $C_\rho, c_\rho, \gamma_\rho$ are $\rho$-dependent constants, monotonically varying in $\rho$,
in ranges
$\pi^2/3<C_\rho<\infty$,
${\pi^2}/{2}<c_\rho<\infty$, 
${\pi^2}/{16}<\gamma_\rho<{\pi^2}/{8}$, 
i.e.,\ 
in between the (ch)GUE and the (ch)G(O,S)E limits.

For a bookkeeping purpose, we exhibit 
plots of the two-level correlation function
$R_2(x,0)$ for nonchiral cases (\ref{Pfaffian}) in Figs.~1 and 2 (right)
and the single-level density $R_1(x)$ for chiral cases in Figs.~3 and 4 (right).
Note that the first peaks of $R_2(x,0)$ and $R_1(x)$ (right figures) are comprised of the corresponding
$P(s)$ and $p_1(s)$ (left figures), respectively.
In the subsequent section,
these analytic results will be tested to fit the Dirac eigenvalue data
numerically obtained from modified SU(2) lattice gauge models.
The practical advantage of adopting distributions of individual level spacings [$P(s)$ and $p_1(s)$]
over $n$-level correlation functions [$R_2(x,0)$ and $R_1(x)$] for fitting is clear from the figures.
As the oscillation of the latter consists of overlapping of multiple peaks,   
the characteristic shape of each peak is inevitably smoothed out, 
leaving us with a rather structureless curve for which an accurate fit is difficult.
On the other hand, the former and its cousins $p_{k}(s)$
(the distribution of the $k^{\rm th}$ smallest eigenvalue \cite{damn})
\be
p_{k}(s)=
-\frac{d}{ds}
\left.\frac{1}{(k-1)!}\(-\frac{\partial}{\partial z}\)^{k-1}\sqrt{{\rm Det}(\openone-z\hat{K}_s)} \right|_{z=1}
\label{k-th}
\ee
that could as well be computed by the Nystr\"{o}m-type method
are very sensitive to the value of the crossover parameter, 
because Eqs.~(\ref{p1s}), (\ref{asympt}) imply that 
the ratio of $P(s)$ or $p_1(s)$ for the orthogonal and sympletic classes to that for the unitary class
grows as $\exp \frac{\pi^2s^2}{16}$ for large $s$ 
(it exceeds $10$ at $s\simeq 2$, $10^2$ at $s\simeq 2.7$, and $10^4$ at $s\simeq 4$).
Therefore $P^{(\rho)}(s)$ and $p^{(\rho)}_1(s)$ should in principle 
admit very sharp one-parameter fitting
by the least square method or simply from the tail of the curve
(in the range of $s$ where the systematic deviation due to finite size is not prominent),
as done for 
the Anderson tight-binding Hamiltonians at the metal-insulator transition \cite{shk}
versus 
the critical random matrix ensembles interpolating G(O,U,S)E and Poisson statistics \cite{nis}.
In the chGSE-chGUE case, the microscopic level density $R_1(x)$ [Fig.~4(b)] would also be suited for fitting,
due to its characteristic oscillatory behavior for a wide range of $x$.

\section{Dirac Spectrum}

\subsection{Antiunitary symmetry}
Dirac operator for the ``real" QCD, i.e., for quark fields belonging to the real or pseudoreal
representation of the gauge group, is known to possess a particular antiunitary symmetry 
unlike that for the complex representation \cite{ver}. 
Namely, Euclidean Dirac operator $D=\gamma_\mu (i\partial_\mu+A_\mu^a\tau_a)$
for quarks in the fundamental representation of SU(2) commutes with $C\tau_{2}K$.
Here, $C$ is the charge conjugation matrix satisfying 
$C\gamma_\mu C^{-1}=-\gamma^\ast_\mu$,
$\tau_2$ is one of the generators of the SU(2) gauge group satisfying
$\tau_2\tau_a \tau_2^{-1}=-\tau_a^\ast$,
and $K$ denotes the complex conjugation.
As $(C\tau_{2}K)^2=+1$, $D$ can be brought to a real symmetric matrix by a similarity transformation.
Similarly, the Dirac operator for quarks in the adjoint representation of SU($N$)
commutes with $CK$.
As $(CK)^2=-1$, $D$ can be brought to a quaternion self-dual matrix.
By the same token, Dirac operators in fundamental representations of O($N$) and Sp(2$N$)
gauge groups are real symmetric and quaternion self-dual, respectively.
It is also well known that
the reality and the self-duality of the continuum Dirac operators are interchanged
for the corresponding Kogut-Susskind staggered Dirac operators [(\ref{DKS}) below], 
i.e., $D^{\rm KS}$ in SU(2) fundamental is essentially quaternion self-dual, whereas
$D^{\rm KS}$ in SU($N$) adjoint is real symmetric, due to the the absence of 
the charge conjugation matrix \cite{hv}.
The spectrum of the SU(2) fundamental staggered Dirac operator was indeed numerically shown to 
belong to the chGSE class \cite{bjmsvw}.

Since the Dirac operator in the fundamental representation of U(1)
(or SU($N'\geq 3$) if one prefers) in continuum or on a lattice  
possesses no such antiunitary symmetry,
the Dirac operator in the
SU(2) fundamental (SU($N$) adjoint)$\times$U(1),
\be
D=\gamma_\mu (i\partial_\mu+ A_\mu^a T_a+B_\mu),
\ee
where $T_a$'s are the generators of the corresponding representation of SU(2) or SU($N$),
and $B_\mu$ the $U(1)$ gauge field, either weakly fluctuating or fixed as a background,
is endowed with a weakly broken
antiunitary symmetry as compared to the pure SU(2) fundamental or SU($N$)-adjoint case.

\subsection{Modified SU(2) lattice gauge theory}
In contrast to the parametric random matrices $H=H_{\rm S}+i\alpha H_{\rm A}$ for which the
breaking of the antiunitary symmetry is uniquely parametrized by
${\displaystyle \alpha=\sqrt{{\<||{\rm Im}\,H||^2\>}/{\<||{\rm Re}\, H||^2\>}}}$,
it is not straightforward to 
identify the bare U(1) gauge coupling constant as proportional to $\alpha$,
due to the gauge invariance:
a Dirac operator that appears complex
could actually be a U(1) gauge transform of some purely real matrix.
In that case, its spectral fluctuation would perfectly be described by chGOE or chGSE.
Thus, in this paper, we restrict ourselves to simpler models without such a subtlety: 
SU(2) quenched lattice gauge theory
associated with twisted boundary condition or coupled with U(1) noise,
each of which does contain an unambiguous counterpart of the $\alpha$ parameter.

We consider SU(2) quenched lattice gauge theory
under the twisted boundary condition (TBC),
that is to multiply  SU(2) links variable at the temporal boundary 
of the hypercubic lattice of size $V=L^4$ by a constant phase\footnote{
As the Dirac operator possesses (pseudo-)reality either for  
periodic or antiperiodic boundary condition, 
we adopt periodic conditions for all directions for simplicity
and consider small deviations from it.}
\be
\e^{i\theta_{n,\mu}}=
\left\{
\begin{array}{ll}
\e^{ i 2\pi  \varphi}& (n_4=L,\ \mu=4)\\
1 & (\mbox{else})
\end{array}
\right. ,
\ee
with $\varphi\ll1$.
This twisting is gauge equivalent to 
the fixed vector U(1) background 
$B_\mu=({2\pi \varphi}/{L})\delta_{\mu,4}$ of AB flux $2\pi\varphi$ \cite{dm}
or  the  imaginary chemical potential $\mu= {i 2\pi \varphi}/{L}$. 
The flux $\varphi$ is the measure of antiunitary-symmetry violation 
and plays the role of 
$\alpha$ in the parametric random matrices. Its effect on the low-energy effective Lagrangian
is completely dictated through the flavor-covariant derivative \cite{svd},
just as in the case of symmetry-violating real chemical potentials \cite{kogut}.
 
We also consider the SU(2)\,+\,U(1) phase noise (PhN) model,
where an independent and identically distributeds random phase $\e^{i\theta_{n,\mu}}$
taken from the uniform distribution $\theta\in [-p,p]\ (p\ll1)$
(one could alternatively adopt Gaussian distribution of variance $p^2$) is multiplied to 
each link SU(2) variable.
The parameter $p$ of randomness is expected to be proportional to 
$\alpha$ in the parametric random matrices,
since
\be
{\frac{\< ||{\rm Im}\,D||^2\>}{\<||{\rm Re}\,D||^2\>}}\simeq
{\frac{\< ||U \sin\theta||^2\>}{\<||U \cos\theta||^2\>}}\simeq \<\theta^2\>=\frac13 p^2.
\ee
The PhN model was previously used to uncover the fake nature \cite{sw} of the 
apparent chGUE-chGSE crossover 
of the Dirac spectrum of Ginsparg-Wilson type \cite{fhlw},
so it would be a good exercise to exploit it again to our
actual universality crossover.

For our purpose of confirming that the lattice model exhibits chG(S,O)E-chGUE
crossover,
it is sufficient to focus on the strong coupling region of SU(2) at $\beta=4/g^{2}=0\sim1.5$, 
where the smoothed level density at the origin, $1/\varDelta(0)$, is sufficiently above zero 
and the chiral symmetry is spontaneously broken,
accepting that the model is away from the continuum limit.
Accordingly, 
we employ the simplest algorithm possible: 
the unimproved plaquette action and
the 10-hit heat-bath update coupled with overrelaxation.
The staggered Dirac operator
\be
D^{\rm KS}_{nm}
=\frac{1}2\(
\eta_{n,\mu} \delta_{n+\hat{\mu},m} U_{n,\mu}\e^{i\theta_{n,\mu}}
-\eta_{m,\mu} \delta_{n, m+\hat{\mu}}U_{m,\mu}^\dagger  \e^{-i\theta_{m,\mu}}
\),\ \ \eta_{n,\mu}\equiv(-1)^{n_1+\ldots+n_{\mu-1}}
\label{DKS}
\ee
is diagonalized using the standard LA package.

Because of our need to detect possibly small deviations of spectral fluctuation
from the universal random matrix statistics at either end ($\rho=0$ or $\infty$),
we give priority to the number of independent samples 
and perform our simulation on a lattice of the smallest size $V=4^4$.
This choice is sufficient for  
measuring the short-distance behavior of eigenvalues within up to 3$\sim$4 mean level spacings
and determining the $\rho$ parameter precisely.
In this region
 the systematic deviation due to the small size of the lattice is less prominent
(they shall manifest at larger separation) than the statistical fluctuation.
Only if one dares to test the symmetry crossover on the large-distance correlation of eigenvalues, 
such as the number variance $\Sigma^2(s)$, would the use of larger lattices become essential.

\subsection{Fitting Dirac spectra}
Our procedure of fitting Dirac spectral statistics to the parametric random matrix predictions
consists of the following steps:
\begin{enumerate}
\item
{\bf Determination of $\varDelta(0)$.}\ 
Perform the pure SU(2) simulation for each $\beta$
and measure the 
fundamental and adjoint staggered Dirac spectrum $\{\lambda_i\}$ for $O(10^5)$ configurations.
Taking for granted that the small Dirac eigenvalues obey the chG(S,O)E statistics, 
determine the mean level spacing at the origin $\varDelta(0)$ in the following two ways:\\
(a) Using the least square method, fit the histogram of the smallest Dirac eigenvalues $\wp_1(\lambda_1)$
to the rescaled chG(S,O)E result (\ref{p1s}), $p_1(\lambda_1/\varDelta)/\varDelta$,
by varying $\varDelta$.
The fitting range is chosen to be within $3\varDelta$ (chGSE) and $2.2\varDelta$ (chGOE),
which is divided into bins of widths $0.1\varDelta$.
Errors of $\varDelta$ are estimated as those increase $\chi^2$ by 1.
\\
(b) Compare the mean value of the smallest Dirac eigenvalue $\<\lambda_1\>$ 
with that of the (unfolded) chiral random matrix eigenvalues from (\ref{p1s}),
\be
\<x_1\>=\int_0^\infty s\,p_1(s)ds=
\sqrt{\frac{2\e}{\pi}}{\rm erfc}\frac{1}{\sqrt{2}}
~~
\mbox{[chGOE]},\ \ 
\sqrt{\frac{2\e}{\pi}}
~~
\mbox{[chGSE]}
\ \ \Rightarrow\ \ \varDelta(0)=\frac{\<\lambda_1\>}{\<x_1\>}.
\ee
Typically, the values of mean level spacing determined by
the above two methods agree within $0.1\%$. 
This provides a ground for a precise determination of the crossover parameter at the origin in step 2.

\item
{\bf Fitting the smallest eigenvalue.}\ Multiply SU(2) link variables $U_{n,\mu}$ by the
phases $\e^{i\theta_{n,\mu}}$ (either TBC or PhN) and
measure the Dirac spectrum $\{\lambda_i\}$ 
for $N_{{\rm conf}}=O(10^{4})$ independent configurations.
The unfolded smallest eigenvalue is still defined by $x_1=\lambda_1/\varDelta(0)$ with
respect to
$\varDelta(0)$ determined in step 1, i.e.~from unmodified SU(2) configurations. 
Fit the frequencies of $x_1$ to the prediction by
the least square method as follows:
we choose the valid range of fitting the smallest eigenvalue $x_1$ 
to be [0, 2.8] (fundamental) and [0, 1.6] (adjoint)
and divide them
into $r=20$ segments ${I_0,\ldots,I_{r-1}}$ of equal width.
The remainder is denoted by $I_r=[$2.8 or 1.6, $\infty)$.
Define
${ 
\chi_\rho^2=\sum_{i=0}^r {(F_i-f_i^{(\rho)})^2}/{f_i^{(\rho)}},
}$
where $F_i$ denotes the frequency of $x_1$ falling on the $i$th interval $I_i$, and
$f_i^{(\rho)}=N_{{\rm conf}} \int_{I_i} p_1^{(\rho)}(s)ds$ its expected value from
the chG(S,O)E-chGUE parametric random matrices.
By varying $\rho$, find the optimal value of $\rho$ that minimizes $\chi^2_\rho$ and 
estimate the error of $\rho$ as that increases $\chi^2_\rho$ by 1.

\item
{\bf Fitting level spacings.}\ 
Fluctuation of Dirac eigenvalues in the spectral bulk is, by itself, not directly related to the
low-energy effective description of the gauge theory but rather to the diffusion of quarks 
in a hypothetical ``time'' evolution with Dirac operator as its Hamiltonian \cite{jnpz2}.
Nevertheless, we shall argue that the distribution of level spacings from the spectral ``plateaux'' 
$[\lambda_{\rm m},\lambda_{\rm M}]$ 
adjacent to, but not including, the origin,
in which the mean level spacing $\varDelta(\lambda)$
is well approximated as a constant close to $\varDelta(0)$ (Fig.5),
provides an efficient way of determining $F$.
It also serves as a clearer indicator of the presence of the universality crossover
in modified SU(2) lattice gauge theories and as a consistency check of the $\rho$ value
directly measured from the smallest eigenvalues distribution.
In order to avoid possible distortion of the level spacing distribution
due to the chiralness, we take $\lambda_{\rm m}$ not too close to the origin
and set it to be, at smallest, the $11^{\rm th}$ eigenvalue.
\\
Define unfolded level spacings in that eigenvalue window as
$s=(\lambda_{i+1}-\lambda_i)/\varDelta(\bar{\lambda}),\ \lambda_i, \lambda_{i+1}\in[\lambda_{\rm m},\lambda_{\rm M}]$.
In case the density profile near the origin is not quite flat (such as
$D^{\rm KS}$ in SU(2) fundamental at $\beta=0.5\sim1.5$), 
we alternatively used unfolding by the linearized level density $\bar{\rho}(\lambda)$, 
$\lambda_i\to x_i=\int^{\lambda_i}\bar{\rho}(\lambda)d\lambda,$ even {\it within} the window chosen,
to offset the explicit local variation in the mean level spacing.
For a window containing $\sim 20$ eigenvalues,
the distortion in $P_{\rm data}(s)$
caused by
the small local inhomogeneity of the $\rho$ parameter through its dependence on $\varDelta=1/\bar{\rho}$
is expected to be negligible.
Find the best fit of the histogram of the unfolded smallest Dirac eigenvalues $P_{\rm data}(s)$
to the crossover chiral random matrix result 
$P^{(\rho)}(s)$
[Figs.~2(a) or 1(a)] by varying $\rho$, as done in step 2.
Thanks to the enormous gain in statistics due to the spectral averaging,
we can safely set the fitting range to be as large as [0, 4] and divide it into
$r=40$ segments.\end{enumerate}
\renewcommand{\thefigure}{{\arabic{figure}}}
\begin{figure}[tb]
\includegraphics[bb=0 0 260 162]{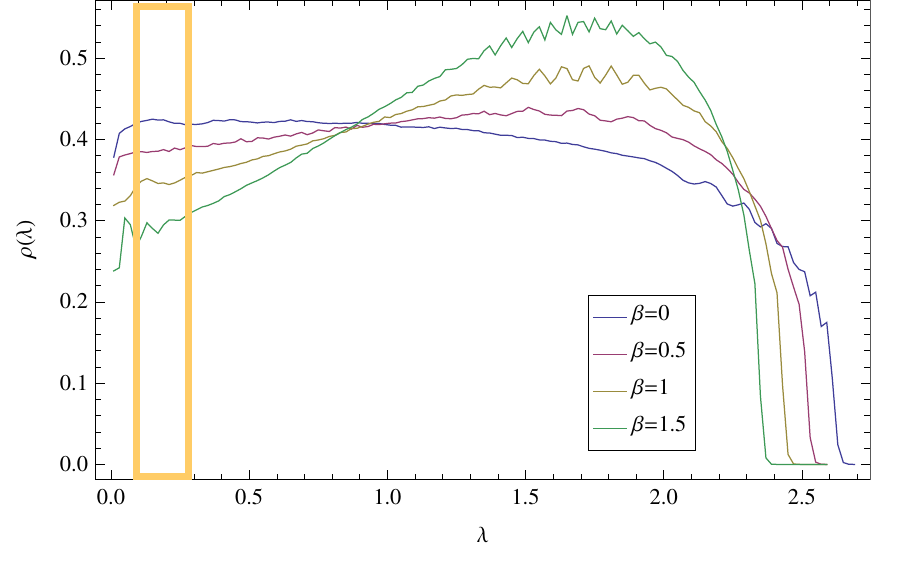}\includegraphics[bb=0 0 260 162]{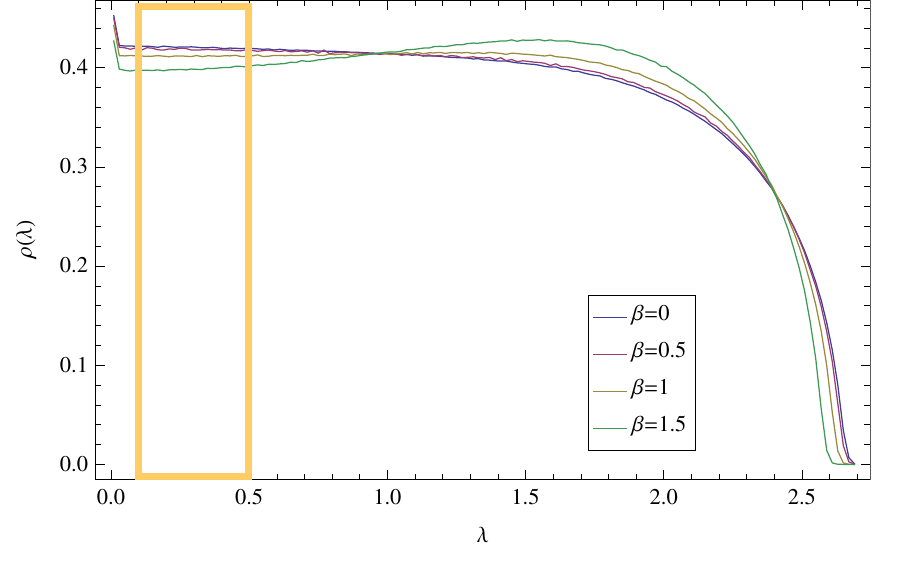}
\caption{Macroscopic level densities $\rho(\lambda)$ of SU(2) fundamental (left) and adjoint (right) $D^{\rm KS}$
at various $\beta$ in the strong coupling region.
Lattice size: $4^4$, number of configurations: $100,000$.
Level spacings from the eigenvalue windows (plateaux) $[\lambda_{\rm m},\lambda_{\rm M}]$ marked by
orange strips in the vicinity of $\lambda=0$ are used for fitting $P^{(\rho)}(s)$.
}
\end{figure}

\subsection{Simulation results}
We present the outcome of one-parameter fitting of
smallest eigenvalue distributions (SED) and level spacing distributions (LSD) to the parametric random matrix ensembles.
We set the symmetry-breaking parameters to be 
$\varphi=0.01\sim.06$ for the SU(2)+TBC model
and
$p=0.02\sim.10$ for the SU(2)+PhN model
and performed simulations at the gauge-coupling constants $\beta=0, 0.5, 1, 1.5$
on a lattice of dimensions $V=4^4$.
$N_{\rm conf}=40,000$ configurations are generated for each set of parameters.
Optimal values of the $\rho$ parameter for the TBC (PhN) model in fundamental (F) and
adjoint (A) representation
are tabulated in the six (seven) columns in the middle of Table I (II).

Sample plots of $p_1(s)$, $R_1(x)$, and $P(s)$ for $\beta=0$ are
exhibited in Figs.6$\sim$8 [SU(2)+TBC] and in Figs.9$\sim$11 [SU(2)+PhN].
In all figures, measured histograms are plotted by colored dots, and parametric
random matrix results that fit the data optimally are shown in curves of the same color
($\varphi=.01$ or $p=.02$ in red to $\varphi=.06$ or $p=.10$ in green).
Also plotted in the figures are the results from chG(S,O)E or G(S,O)E (black real line),
and chGUE or GUE (broken line).
Note that the microscopic level density $R_1(x)$ is not fitted to the corresponding data
by the least square method; 
the $\rho$ parameter determined from the SED is adopted to the respective $R_1(x)$.

Even at inspection, one is convinced of the accuracy of one-parameter fitting in all cases.
The $\chi^2$/d.o.f.\ deviations from optimally fitting random matrix distributions
are summarized in Table III.
Considering the smallness of our lattice ($4^4$),
the goodness-of-fit achieved is astonishing
(except for LSDs for SU(2) fundamental at very small $\rho\lesssim 0.1$,
where the distribution becomes extremely peaky at small $s$ due to the onset of
Kramers degeneracy and the fitting error is inevitably enhanced).
These listed values are comparable to those reported in the pioneering papers in this field \cite{dam1} 
using chGUE-chGUE (i.e.\ Hermitian) crossover,
fitted to the spectral data from larger lattices: 
$\chi^2$/d.o.f.$=0.33$ for quenched QCD on $12^4$ and
$\chi^2$/d.o.f.$=1.13\sim1.33$ for dynamical QCD on $6^4$.

From Tables I and II one immediately notices that
for a fixed bare coupling $\beta$, $\varphi$-$\rho$ and $p$-$\rho$ plots are all quite linear. 
In Fig.~12 we exhibit sample plots of
${\sqrt{\varDelta}}\rho/\varphi$ for SU(2)+TBC and ${\sqrt{\varDelta}}\rho/p$ for SU(2)+PhN \footnote{%
The factor ${\sqrt{\varDelta}}$ is included to facilitate extraction of $F^2/\Sigma$ in the next section.},
showing that
(i) these ratios, namely from LSDs, are very stable
under the change of symmetry-violating parameters $\varphi$ or $p$
and
(ii) the values of $\rho$ determined from SED and from LSD are in a good agreement, as should be.
This precisely linear dependence is essential in accurately determining the pion decay constant
from SU(2)+TBC models in the next section.
We note that in the strong coupling limit $\beta=0$,
four plots
nearly overlap on top of each other,
regardless of the representation being fundamental (F) or adjoint (A)
(see also Fig.~13).
The splitting between fundamental and adjoint becomes more apparent at larger value of $\beta$.

\begin{table}
\centering
\caption{SU(2) + TBC model on $V=4^4$: Crossover parameters and low-energy constants}\label{table1}
\begin{tabular}{|c|c||c|c|c|c|c|c|c|c|c|c|c|c|}
\hline
\hline
$\beta$&rep/dist&&$\varDelta$&$\Sigma$&
\multicolumn{6}{|c|}{$\rho$}
&$\sqrt{\varDelta}\rho/\mu$&$F^2/\Sigma$&$F^2$ \\ \cline{6-11}
&&
$[\lambda_{\rm m},\lambda_{\rm M}]$
&
&&$\varphi$=.01&.02&.03&.04&.05&.06&&&\\
\hline
\hline
0& F/SED &            &.00933(1)&1.315(2)&.061(1)&.123(1)&.181(1)&.245(2)&.305(2) &.364(2) &.375(1)&.221(1)&.290(2) \\
 & F/LSD & [.10, .30] &.00929(0)&1.321(0)&.061(0)&.122(0)&.183(1)&.243(1)&.303(2) &.364(4) &.374(1)&.219(1)&.290(1) \\ \cline{2-14}
 & A/SED &            &.00617(0)&1.989(1)&.076(1)&.153(2)&.224(3)&.309(5)&.391(8) &.453(11)&.382(2)&.229(3)&.455(6) \\
 & A/LSD & [.10, .50] &.00618(0)&1.984(0)&.075(2)&.150(4)&.224(6)&.297(8)&.371(14)&.447(25)&.374(5)&.219(5)&.435(11)\\ \cline{2-14}
 \hline
0.5&F/SED&            &.01020(1)&1.203(2)&.055(1)&.110(1)&.160(1)&.210(2)&.263(2) &.313(2) &.339(1)&.180(1)&.217(1) \\
 & F/LSD & [.10, .26] &.01004(0)&1.223(0)&.052(0)&.103(0)&.155(1)&.206(1)&.257(2) &.306(2) &.329(1)&.170(1)&.208(1) \\ \cline{2-14}
 & A/SED &            &.00626(2)&1.960(6)&.067(1)&.131(2)&.200(3)&.263(4)&.327(6) &.386(8) &.332(2)&.173(2)&.339(5) \\
 & A/LSD & [.10, .50] &.00622(0)&1.972(0)&.067(3)&.134(4)&.202(5)&.269(7)&.333(11)&.407(17)&.337(4)&.179(5)&.353(9) \\ \cline{2-14}
 \hline
1.0&F/SED&            &.01150(5)&1.067(5)&.056(1)&.102(1)&.148(1)&.193(1)&.241(2) &.286(2) &.329(1)&.170(1)&.181(1) \\
 & F/LSD & [.10, .30] &.01105(0)&1.110(0)&.048(0)&.095(0)&.143(1)&.190(1)&.237(1) &.284(2) &.319(1)&.160(1)&.177(1) \\ \cline{2-14}
 & A/SED &            &.00631(3)&1.944(9)&.065(1)&.132(2)&.200(3)&.266(4)&.338(6) &.397(8) &.336(2)&.177(2)&.344(5) \\
 & A/LSD & [.10, .50] &.00632(0)&1.943(0)&.066(3)&.133(4)&.199(5)&.266(7)&.331(10)&.394(16)&.335(4)&.176(4)&.343(9) \\ \cline{2-14}
 \hline
1.5&F/SED&            &.01412(1)&0.869(0)&.061(1)&.101(2)&.138(1)&.178(1)&.218(2) &.257(2) &.329(1)&.170(1)&.148(1) \\
 & F/LSD & [.13, .28] &.01280(0)&0.959(0)&.043(0)&.085(0)&.127(1)&.169(1)&.210(1) &.251(1) &.319(1)&.160(1)&.153(1) \\ \cline{2-14}
 & A/SED &            &.00654(1)&1.877(3)&.067(1)&.128(2)&.193(3)&.251(4)&.312(6) &.392(8) &.336(2)&.177(2)&.333(5) \\
 & A/LSD & [.10, .50] &.00650(0)&1.887(0)&.065(3)&.128(4)&.194(5)&.257(7)&.322(10)&.386(15)&.335(4)&.176(4)&.333(8) \\ \cline{2-14}
\hline
\hline
\end{tabular}
\end{table}
\begin{table}
\centering
\caption{SU(2) + PhN model on $V=4^4$: Crossover parameters}\label{table2}
\begin{tabular}{|c|c||c|c|c|c|c|c|c|c|c|c|}
\hline
\hline
$\beta$&rep/dist&&$\varDelta$&
\multicolumn{7}{|c|}{$\rho$}
&$\sqrt{\varDelta}\rho/p$  \\ \cline{5-11}
&&$[\lambda_{\rm m},\lambda_{\rm M}]$
&
&
$p$=.02&.03&.04&.05&.06&.08&.10& \\
\hline
\hline
0  & F/SED &            &.00933(1)&.079(1)&.117(1)&.154(1)&.194(1)&.235(2)&.311(2) &.390(1) &.376(0) \\
   & F/LSD & [.10, .30] &.00929(0)&.078(0)&.117(0)&.156(1)&.194(1)&.233(1)&.311(2) &.387(3) &.375(0) \\ \cline{2-12}
   & A/SED &            &.00617(0)&.094(2)&.145(2)&.193(3)&.244(4)&.288(5)&.381(8) &.481(7) &.377(2) \\
   & A/LSD & [.10, .50] &.00618(0)&.095(3)&.142(4)&.191(5)&.237(6)&.285(8)&.383(15)&.471(18)&.374(4) \\ \cline{2-12}
 \hline
0.5& F/SED &            &.01020(1)&.071(1)&.106(1)&.143(1)&.178(1)&.213(1)&.286(1) &.353(2) &.360(0) \\
   & F/LSD & [.10, .26] &.01009(0)&.072(0)&.108(0)&.143(1)&.179(1)&.215(1)&.286(1) &.359(4) &.361(0) \\ \cline{2-12}
   & A/SED &            &.00626(2)&.094(2)&.140(2)&.185(3)&.230(4)&.274(4)&.367(5) &.446(10)&.365(2) \\
   & A/LSD & [.10, .50] &.00622(0)&.094(3)&.142(4)&.189(5)&.235(6)&.283(8)&.377(10)&.468(28)&.372(4) \\ \cline{2-12}
 \hline
1.0& F/SED &            &.01150(5)&.062(1)&.094(1)&.125(1)&.156(1)&.188(1)&.255(2) &.316(2) &.338(0) \\
   & F/LSD & [.10, .30] &.01120(0)&.065(0)&.097(0)&.130(0)&.162(1)&.194(1)&.259(2) &.322(3) &.343(0) \\ \cline{2-12}
   & A/SED &            &.00631(3)&.092(2)&.140(2)&.188(3)&.233(4)&.286(5)&.376(7) &.490(12)&.373(2) \\
   & A/LSD & [.10, .50] &.00632(0)&.094(3)&.138(4)&.186(5)&.231(6)&.280(8)&.374(14)&.459(27)&.369(4) \\ \cline{2-12}
\hline
\hline
\end{tabular}
\end{table}
\begin{table}
\centering
\caption{$\chi^2/$d.o.f. for fitting spectral distributions}
\begin{tabular}{|c|c||c|c|c|c|c|c|c|}
\hline
\hline
dist&rep&\multicolumn{4}{|c|}{SU(2)+TBC}&\multicolumn{3}{|c|}{SU(2)+PhN}\\ \cline{3-9}
&&$\beta$=0&0.5&1.0&1.5&$\beta$=0&0.5&1.0 \\ \hline\hline
SED&F&0.79\s1.22 & 0.54\s1.00 & 0.64\s1.80 & 0.86\s3.13 & 0.58\s1.33 & 0.63\s1.50 & 0.73\s1.44 \\ \cline{2-9}
   &A&0.61\s1.11 & 0.36\s1.24 & 0.80\s1.59 & 0.79\s1.31 & 0.41\s1.42 & 0.65\s1.88 & 0.60\s1.49 \\ \hline
LSD&~F*&0.65\s1.07 & 1.11\s1.53 & 0.63\s1.61 & 0.89\s1.41 & 0.61\s1.54 & 0.55\s1.29 & 0.78\s1.71 \\ \cline{2-9}
   &A&0.97\s1.55 & 0.63\s1.09 & 0.57\s1.15 & 0.60\s1.17 & 0.61\s1.40 & 0.80\s1.22 & 0.72\s1.21 \\ \hline\hline
\end{tabular}\\
$*$ excludes F/LSD for TBC at $\rho\lesssim0.1$, for which $\chi^2/$d.o.f.=$3\sim5$.
~~~~~~~
\end{table}

\begin{figure}
\includegraphics[bb=00 00 260 169]{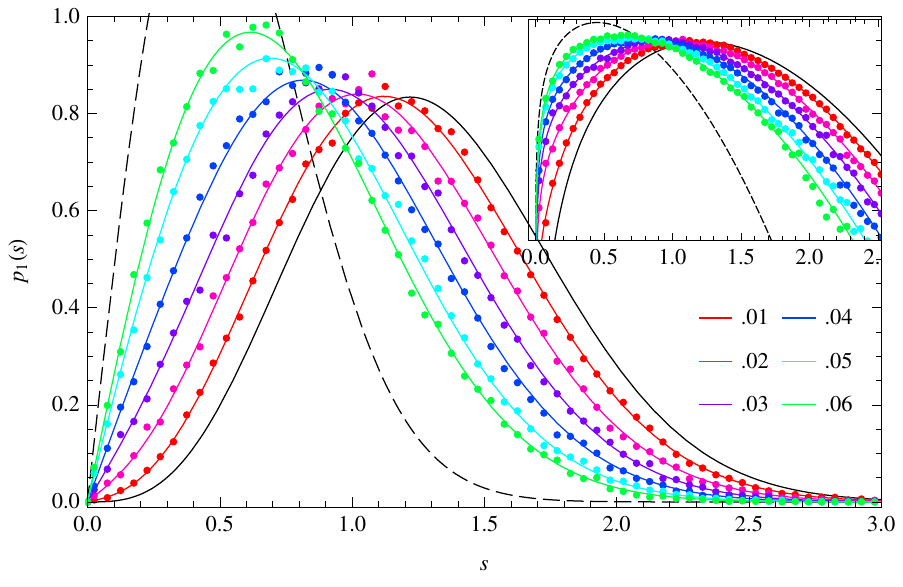}~\includegraphics[bb=00 00 259 164]{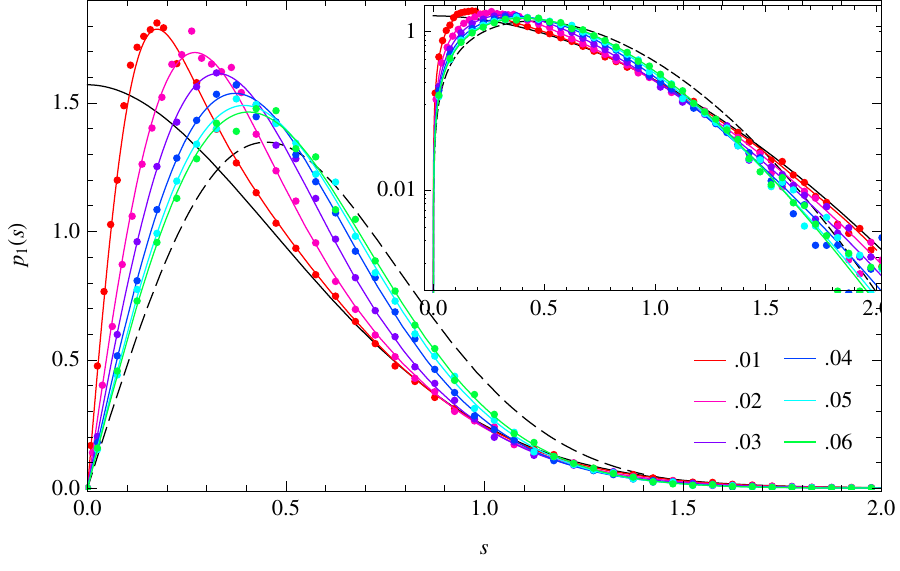}
\caption{
Smallest eigenvalue distributions of SU(2)+TBC $D^{\rm KS}$ in fundamental (left) and adjoint (right),
at $\beta=0$ and $\varphi=0.01\sim.06$. 
}
\vspace{5mm}
\includegraphics[bb=00 00 260 167]{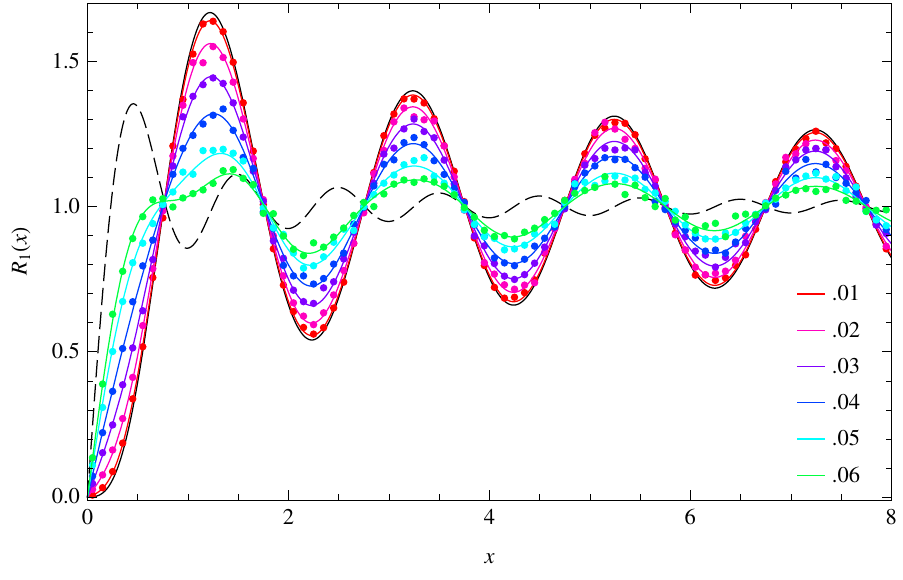}~\includegraphics[bb=00 00 259 163]{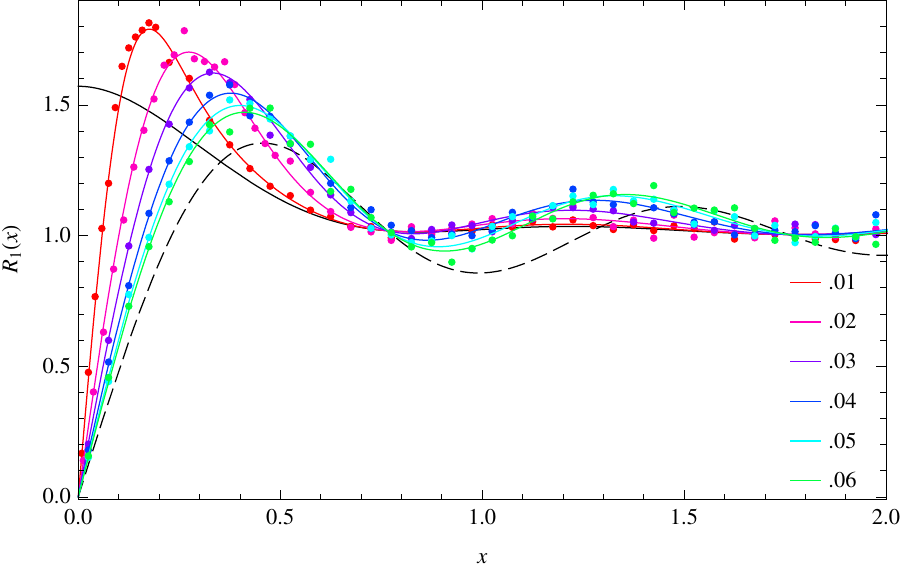}
\caption{
Microscopic level densities of SU(2)+TBC $D^{\rm KS}$ in fundamental (left) and adjoint (right),
at $\beta=0$ and $\varphi=0.01\sim.06$. 
}
\vspace{5mm}
\includegraphics[bb=00 00 260 168]{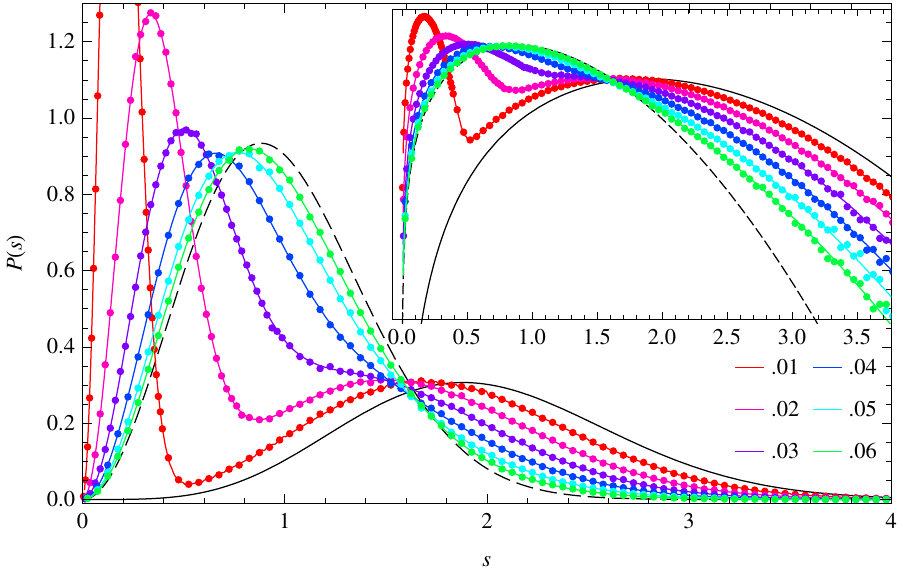}~\includegraphics[bb=00 00 260 171]{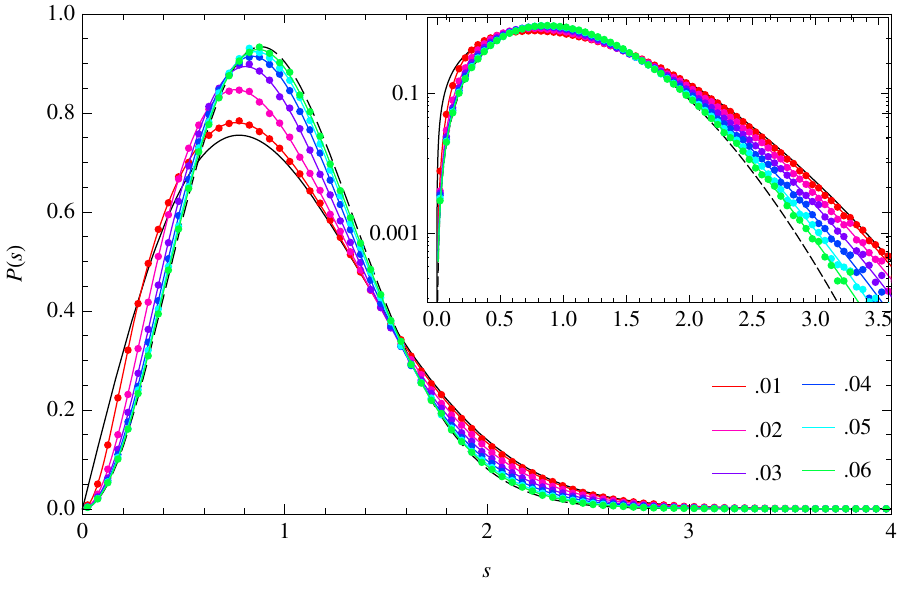}
\caption{
Level spacing distributions of SU(2)+TBC $D^{\rm KS}$ in fundamental (left) and adjoint (right),
at $\beta=0$ and $\varphi=0.01\sim.06$. 
}
\end{figure}
\begin{figure}
\includegraphics[bb=00 00 259 164]{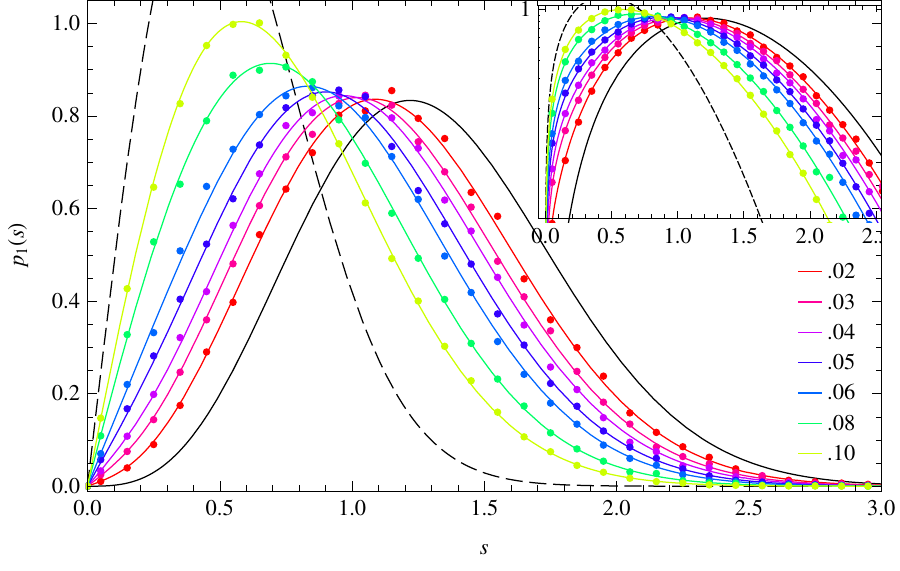}~\includegraphics[bb=00 00 259 164]{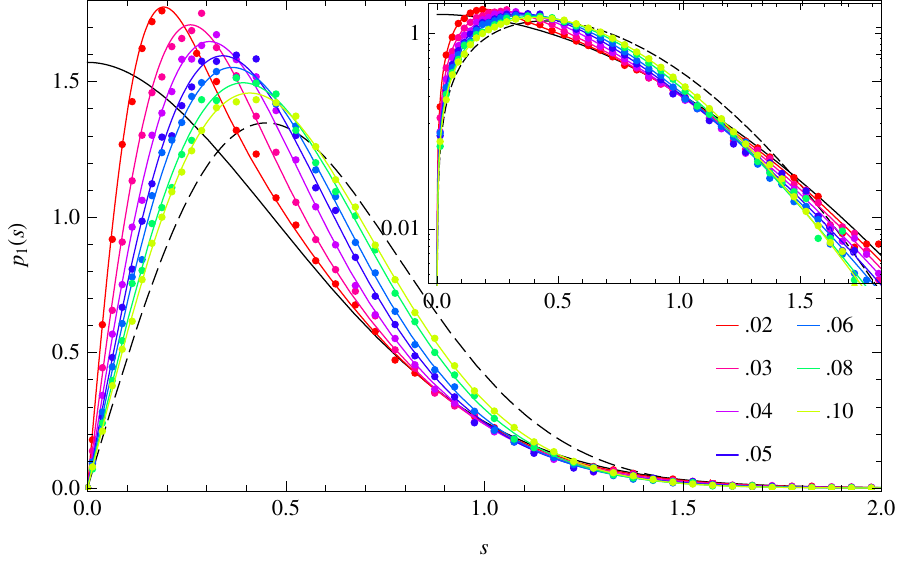}
\caption{
Smallest eigenvalue distributions of SU(2)+PhN $D^{\rm KS}$ in fundamental (left) and adjoint (right),
at $\beta=0$ and $p=0.02\sim.10$. 
}
\vspace{5mm}
\includegraphics[bb=00 00 260 167]{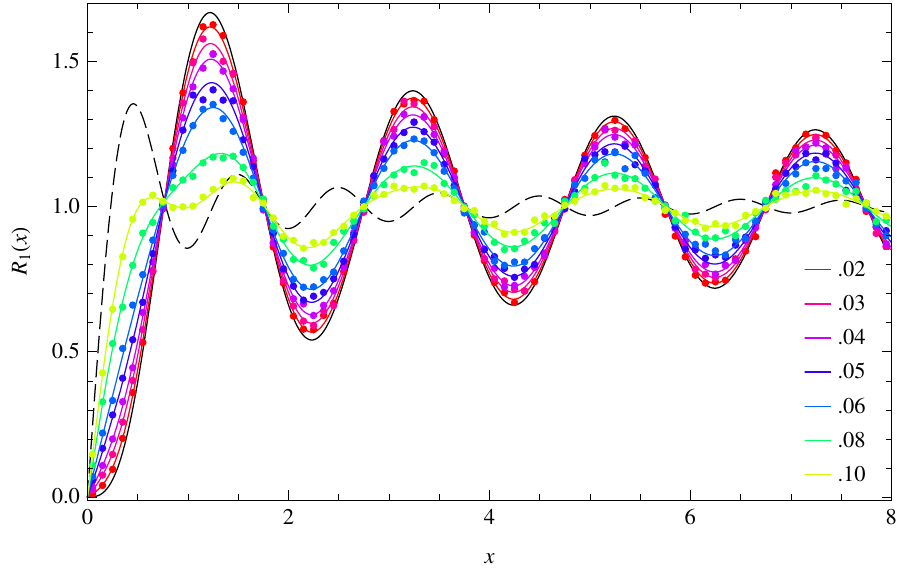}~\includegraphics[bb=00 00 259 163]{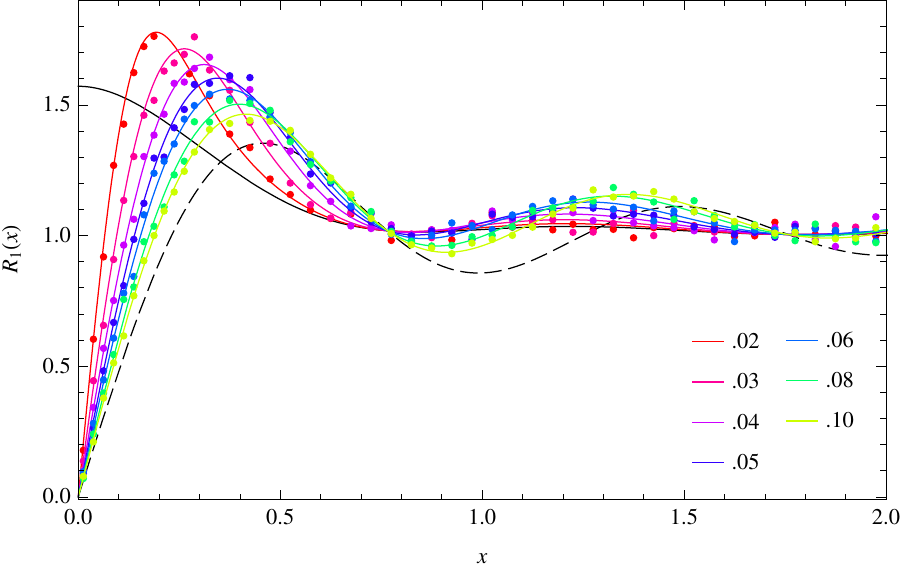}
\caption{
Microscopic level densities of SU(2)+PhN $D^{\rm KS}$ in fundamental (left) and adjoint (right),
at $\beta=0$ and $p=0.02\sim.10$. 
}
\vspace{5mm}
\includegraphics[bb=00 00 260 168]{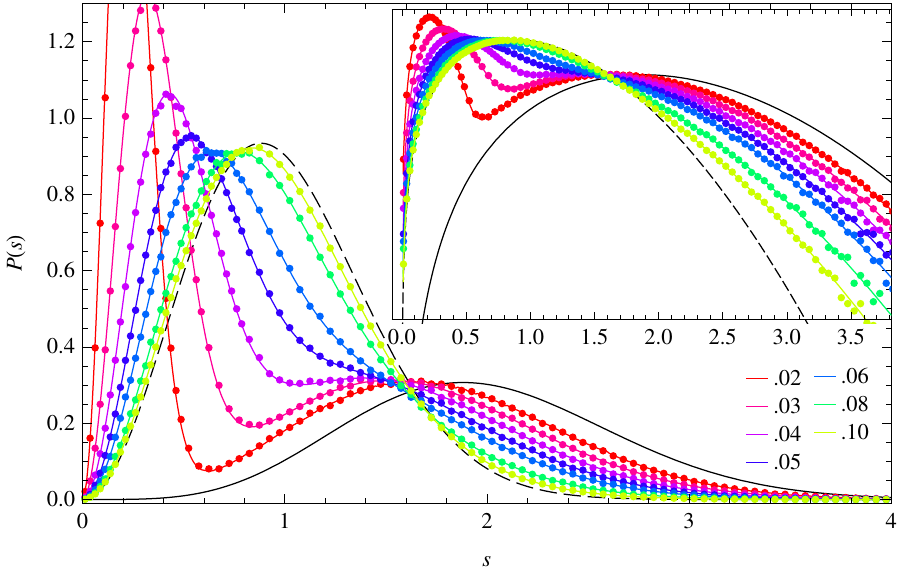}~\includegraphics[bb=00 00 260 170]{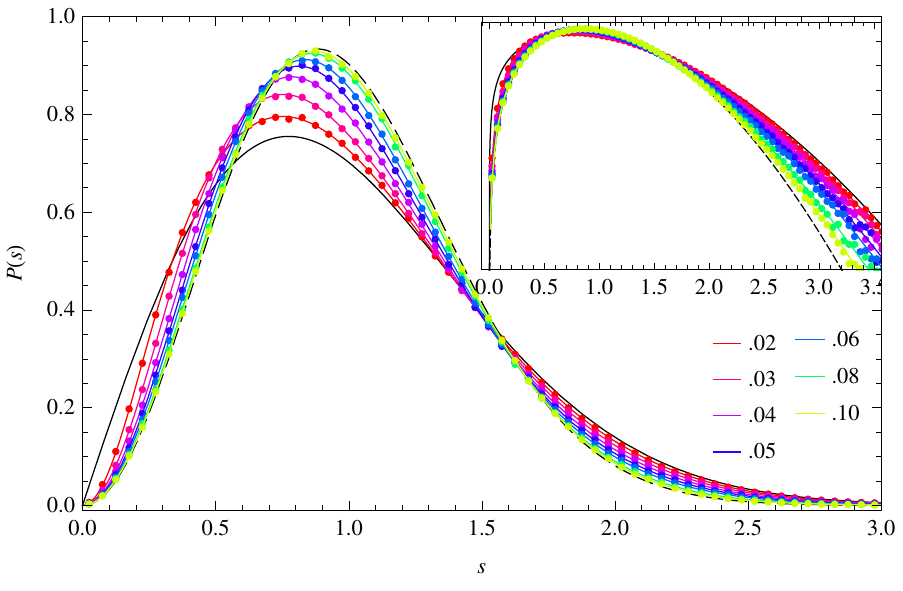}
\caption{
Level spacing distributions of SU(2)+PhN $D^{\rm KS}$ in fundamental (left) and adjoint (right),
at $\beta=0$ and $\varphi=0.02\sim.10$. 
}
\end{figure}

\begin{figure}[t]
\includegraphics[bb=0 0 244 158]{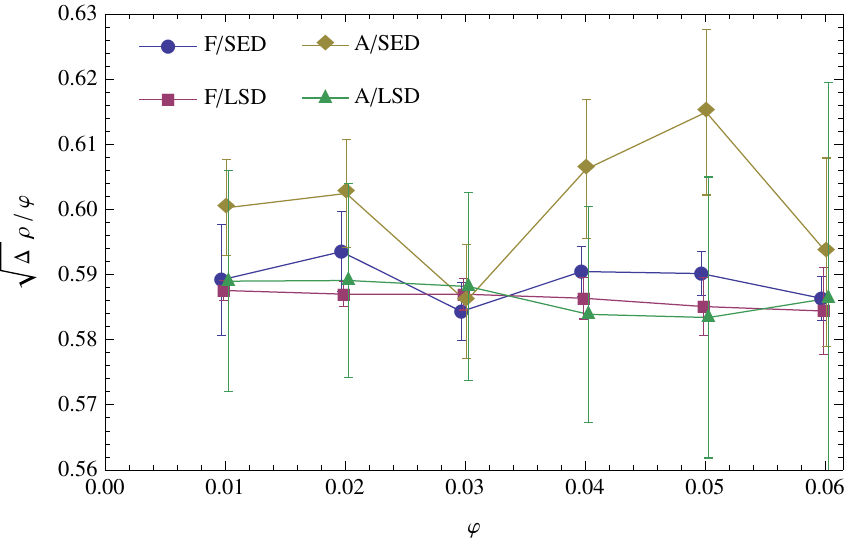}~~\includegraphics[bb=0 0 244 155]{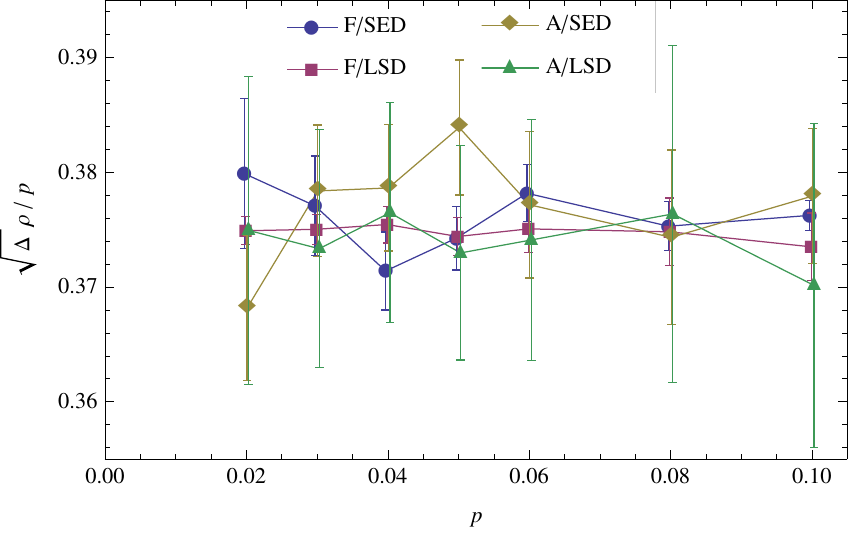}
\caption{The ratio between
the symmetry-violating parameters $\varphi$ and $p$ versus 
the crossover parameter $\sqrt{\varDelta}\rho$ for 
SU(2)+TBC model (left) and SU(2) + PhN model (right) in the strong coupling limit $\beta=0$,
as tabulated in Table I and II.
}
\end{figure}

\section{Low-Energy Constants}
\subsection{Chiral Lagrangian}
Below, we briefly review the chiral Lagrangian description of
QCD-like theories and sketch
the identification of the low-energy constants with the parameters of random matrices.
The effective low-energy Lagrangian for QCD-like theories with $N_F$ flavors of quarks
in (pseudo-)real representation,
at finite chemical potential $\mu$ and bare quark mass $m$
is unambiguously fixed by the global symmetry alone
(provided that $\mu$ is much smaller than the $\rho$ meson mass)
and takes the form containing
two phenomenological free parameters $F$ and $\Sigma$
\cite{kogut},
\be
{\cal L}_{\rm eff}(Q)=\frac{1}{2} F^2\,{\rm tr}\, \partial_\mu Q^\dagger  \partial_\mu Q
+2 F^2\mu \,{\rm tr}\, \hat{B} Q^\dagger \partial_0 Q
-F^2\mu^2 \,{\rm tr}\, (\hat{B} Q^\dagger\hat{B} Q+\hat{B}\hat{B})
-\frac{1}{2} \Sigma m\,{\rm Re}\,{\rm tr}\, \hat{M} Q.
\ee
Here $Q(x)$ is an SU($2N_F$) matrix-valued Nambu-Goldstone field,
$\hat{B}=\sigma_3 \otimes \openone_{N_F}$, 
$\hat{M}=\sigma_1\otimes \openone_{N_F}$\ $\(i\sigma_2 \otimes \openone_{N_F}\)$ 
for quarks in a real (pseudoreal) representation.
$F$ is the ``pion'' decay constant and $\Sigma=\<\bar{\psi}\psi\>/N_F$ the chiral condensate,
both measured in the chiral and zero-chemical potential limit $m,\, \mu\to 0$.
If the theory is in a finite volume $V=L^4$ and
Thouless energy defined as
${E_c\simeq {F^2}/{\Sigma L^2}}$
is much larger than $m$, 
the path integral is dominated by the zero-mode integration
\be
Z=\int_{{\rm SU}(2N_F)} \!\!\!\!\!\!\!\!\!\!\!\!\! dQ\,\,\,\exp\(
V\mu^2 F^2\,{\rm tr}\, (\hat{B} Q^\dagger\hat{B} Q+\hat{B}\hat{B})
+\frac12 V\Sigma m \,{\rm Re}\,{\rm tr}\, \hat{M} Q
\),
\label{Zchiral}
\ee
and the theory is said to be in the $\varepsilon$ regime.
In order to extract the Dirac spectrum,
one introduces fictitious bosonic quarks as well as fermionic quarks in the fundamental theory,
leading to the graded group version of (\ref{Zchiral}) on the effective theory side.
For the actual computation,
one needs to parametrize the graded matrix $Q$ in terms of its eigenvalues.
Comparing the resulting expression
(after analytic continuation $\mu\to i\mu$ and $m\to i\lambda$)
with the random matrix results (\ref{chGOEchGUE}) and (\ref{chGSEchGUE}),
the coefficients of chemical-potential and ``mass'' terms in the exponents on both sides
are readily identified as
$4VF^2\mu^2=2\pi^2\rho^2$ and $V\Sigma \lambda=\pi x.$
The latter is merely the definition of unfolded eigenvalues $x=\lambda/\varDelta$,
when combined with Banks-Casher relation $\Sigma=\pi/\varDelta V$ that determines
one of the low-energy constants $\Sigma$ in terms of the mean spacing $\varDelta$ of small Dirac eigenvalues.
Eliminating the volume in favor of the level spacing, the former equation reads
\be
{\sqrt{\varDelta}}\,\rho=\sqrt{\frac{2 }{\pi}\frac{F^2}{\Sigma }} \mu,
\ee
where the left-hand side is a volume-independent combination.
Accordingly, one can determine another low-energy constant
$F^2/\Sigma$ from the slope of
$\mu$-${\sqrt{\varDelta}}\rho$ plots, preferably on lattices of various sizes.
Note that in the parameter region $V\Sigma |m|\gg1$, Eq.(\ref{Zchiral}) should
approach the $\sigma$ model of nonchiral parametric random matrix ensembles \cite{aie}, 
but we see no reason why
the ``pion decay'' or diffusion constant multiplying ${\rm tr}\, \hat{B} Q^\dagger\hat{B} Q$ is affected.
Accordingly, if the mean level spacing is approximately constant in a window in the very vicinity of the origin,
one can determine $F^2/\Sigma$ from the {\it bulk} correlation (namely the LSD)
in that window.

\subsection{SU(2) + twisted BC}
As the imaginary chemical potential is equal to the flux per link, $\mu=(2\pi/L)\varphi$,
$F^2/\Sigma$ can be extracted from the
$\varphi$-${\sqrt{\varDelta}}\rho$ plots for the SU(2)+TBC model.
In Table I we exhibit the values of
$F^2/\Sigma$ determined from the slopes of these $\varphi$-${\sqrt{\varDelta}}\rho$ plots,
with all the numerals in the lattice unit.
We also list the mean level spacings ($\varDelta(0)$ for the SED rows
and $\varDelta([\lambda_{\rm m},\lambda_{\rm M}])$ for the LSD rows)
and the chiral condensate $\Sigma$.
The pion decay constant $F^2$ is then obtained by multiplying $F^2/\Sigma$ and $\Sigma$
in each row.
Coupling dependence of these low-energy constants as determined from
respective spectral distributions
are summarized in Fig.13.
In order to offset the number of components in the gauge multiplet,
the constants for adjoint quarks are multiplied by 2/3 in the plot.
We notice that
except for the SU(2) fundamental at $\beta=0.5$ and 1.5,
two values of $F^2$, one determined solely from the zero-virtuality correlations (SED) and the
other from the bulk correlations (LSD), agree within the one-$\sigma$ error bars.
Although it is admittedly difficult to calibrate the systematic deviations
involved in employing the bulk correlation for the determination of $F$
(part of which could originate from nonzero modes of the effective theory
or from the nonuniformity of the mean level spacing within the eigenvalue window (see Fig.~5, left)
due to the smallness of the lattice),
these numerical agreements could justify the latter method using LSDs {\it a posteriori}.
\begin{figure}[t]
\includegraphics[bb=00 00 108 120]{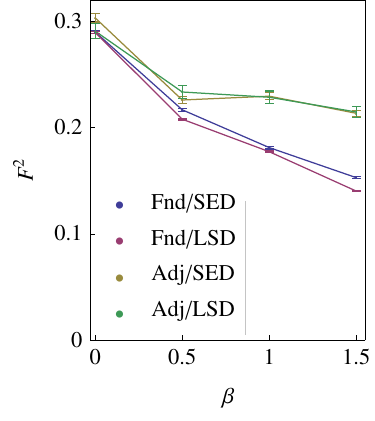}~~\includegraphics[bb=00 00 108 127]{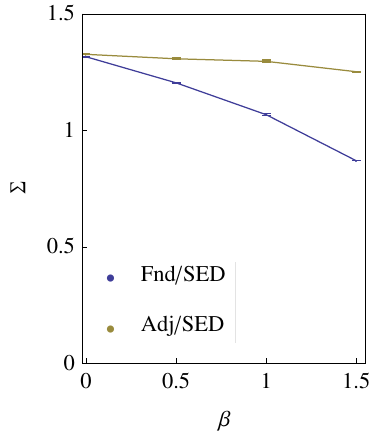}
\caption{
Low-energy constants $F^2$ and $\Sigma$ for SU(2) quenched lattice gauge theory.
Values are multiplied by 2/3 for the adjoint. 
}
\end{figure}

At $\beta=0$, both low-energy constants agree between fundamental and adjoint representations.
This observation is consistent with the fact that in the strong coupling limit 
the low-energy constants are identical to the Sp($N$) and O($N$) lattice gauge theories
at large $N$ \cite{nn},
which share the same antiunitary symmetries as SU(2) fundamental and adjoint.
We also notice that due to the onset of chiral symmetry restoration at $\beta=1.5$, 
the microscopic level density $R_1(x)$ for SU(2) fundamental
starts to deviate from the random matrix result at $x\gtrsim 2$ (Fig.~14, left).
Thus the range of $x$ available for fitting the SED
is rather limited, especially on a lattice as small as $V=4^4$.
On the other hand, the LSDs near the origin, not being directly sensitive to the chiral symmetry,
are still fittable to the random matrix result without noticeable deviation
throughout the plotted region (Fig.~14, right),
and the values of $\sqrt{\Delta}\rho$ determined from the LSD retain the linear dependence on $\varphi$.
Therefore, we consider
the fitting of LSDs to the parametric nonchiral random matrices
to be an efficient method of extracting $F^2$ from small lattices.
\begin{figure}[h]
\includegraphics[bb=0 0 259 166]{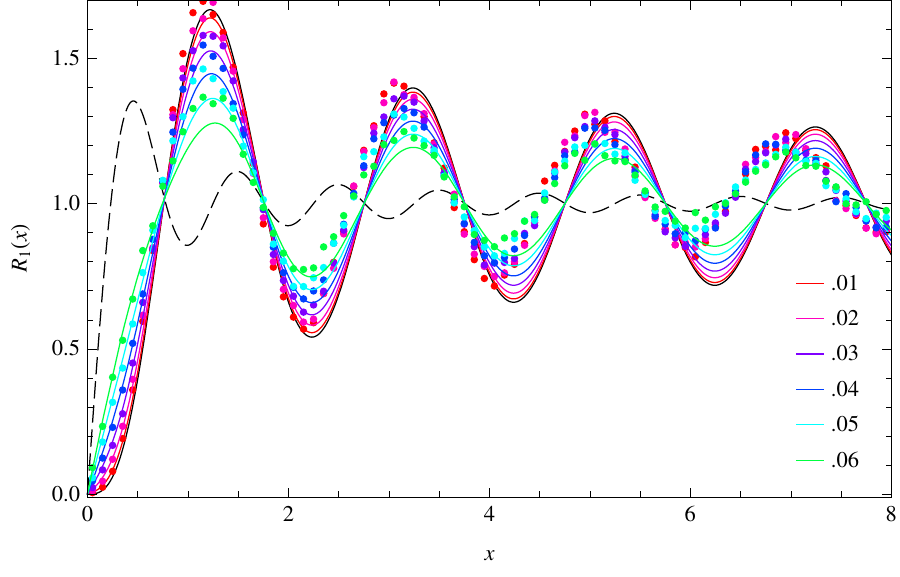}~\includegraphics[bb=0 0 259 170]{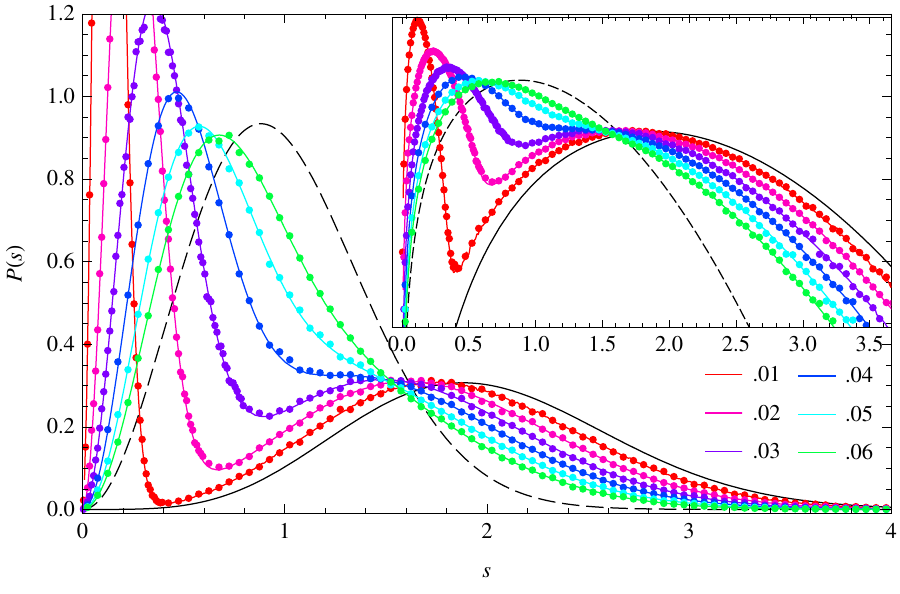}
\caption{
Microscopic level density (left) and level spacing distribution (right)
of SU(2)+TBC $D^{\rm KS}$ in fundamental representation,
at $\beta=1.5$ and $\varphi=0.01\sim.06$.
}
\end{figure}

\subsection{SU(2) + phase noise}
The strength of phase randomness $p$ in the SU(2)+PhN model is not related to the
imaginary chemical potential and thus cannot be directly used to determine $F^2$
from the response of the Dirac spectra.
Thus, we merely list in Table II the slopes of $p$-${\sqrt{\varDelta}}\rho$ plots as a counterpart of 
$\mu$-${\sqrt{\varDelta}}\rho$ plots.
We, however, noticed, rather unexpectedly, that for the SU(2) theory at the strong coupling limit $\beta=0$,
the ``conversion ratio'' between the phase randomness $p$ and the imaginary chemical potential $\mu$
is unity within numerical error
(see the $\sqrt{\varDelta}\rho/\mu$ column in Table I and the $\sqrt{\varDelta}\rho/p$ column in Table II).
This fact will have to be accounted for analytically.
They gradually disagree for increasing $\beta$.

Finally, we need to check the volume dependence of the crossover parameter, which should scale as
$\rho=(\sqrt{2}F\mu/\pi)\sqrt{V}$.
Only for the purpose of varying the volume of the lattice in small steps,
we adopted the two-dimensional toy model of SU(2)+PhN at $\beta=0$.
$N_{\rm conf}=22,000$ independent configurations are generated for each set of parameters.
First, we confirmed that the model shares the features in four dimensions,
i.e., all spectral distributions fit nicely to the parametric random matrices,
and $\sqrt{\Delta}\rho$ scales with the randomness $p$
on a lattice of fixed size $V=16^2$ (Fig.15).
The eigenvalue windows for sampling levels spacings
are chosen to be $[\lambda_{\rm m}, \lambda_{\rm M}]=[0.30,0.65]$ (fundamental) 
and [0.16, 0.60] (adjoint).
Then the $\rho$ parameter is measured from the level spacing distributions
on lattices of size $V=10^2\sim24^2$ at fixed phase randomness $p=0.03$.
$\rho$ is indeed seen to scale as expected, linearly increasing in $\sqrt{V}$ (Fig.16).
\begin{figure}
\includegraphics[bb=0 0 259 161]{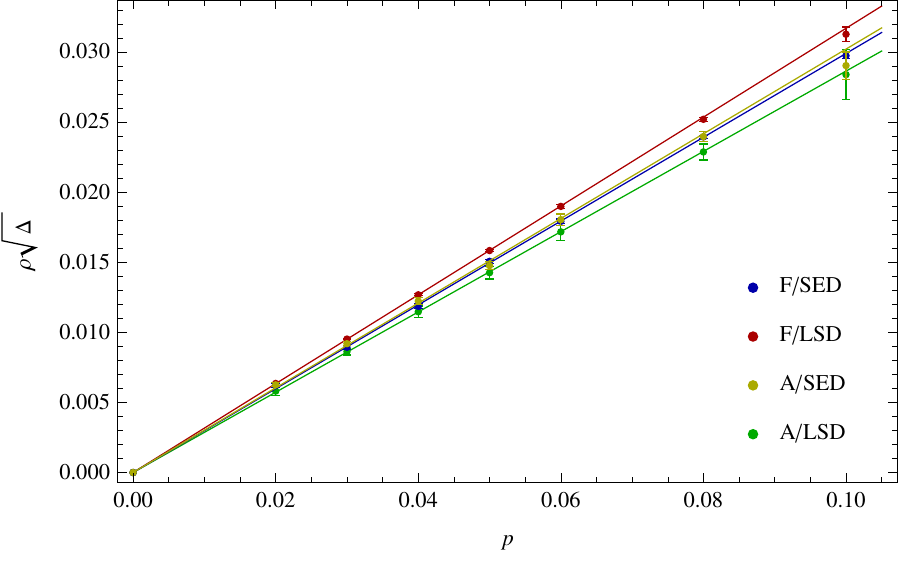}~\includegraphics[bb=0 0 259 170]{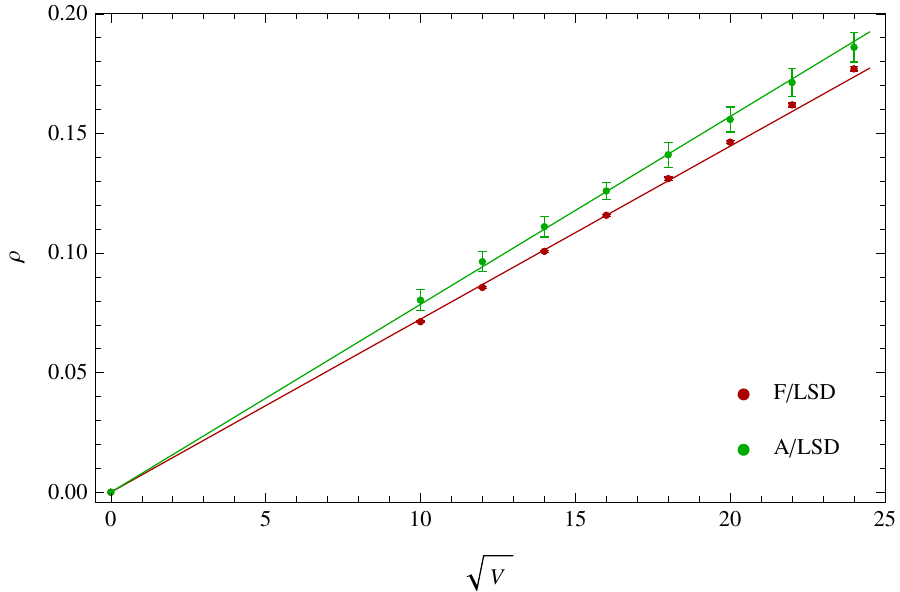}
\caption{(left) Crossover parameter $\sqrt{\Delta}\rho$ for
SU(2)+PhN model at $\beta=0$ and $p=0.02\sim.10$.
Lattice size: $16^2$, number of configurations: $22000$.
Eigenvalue windows: $[\lambda_{\rm m}, \lambda_{\rm M}]=[0.30,0.65]$ (F), [0.16, 0.60] (A).
Real lines are linear fits to the data.
}
\caption{(right) Volume dependence of the crossover parameter $\rho$ for
SU(2)+PhN model at $\beta=0$ and $p=0.03$, determined from LSDs.
Lattice size: $10^2\sim24^2$. Other parameters are common to Fig.15.
}
\end{figure}

\section{Conclusions}
In this paper we have computed the 
level spacing and smallest eigenvalue distributions
of Hermitian random matrices in the G(O,S)E-GUE and chG(O,S)E-chGUE crossover.
The results are shown to be
perfectly fittable to the adjoint and fundamental staggered Dirac spectra of
quenched SU(2) lattice gauge theory whose antiunitary symmetry is weakly violated by
twisted boundary condition or noisy phases.
This leads to the precise determination of the pion decay constant $F$
(in the chiral and zero-density limit)
from the Dirac spectral data.
This method, feasible on a small-size lattice, has an advantage over the conventional method
of measuring the decay rate of axial correlators, which requires a large temporal dimension.

Our treatment is complementary to the previous approach of determining $F$ of
two-color QCD from its Dirac spectrum: 
Akemann and collaborators have concentrated on the {\it real} chemical potential,
measured the response of Dirac eigenvalues that permeate into the complex plane,
and fitted the lattice data to the {\it non-Hermitian} parametric chiral random matrices \cite{ake}.
On the other hand, we considered the twisted boundary condition, i.e., the
{\it imaginary} chemical potential and
measured the response of Dirac eigenvalues that cross over within the real axis,
and fitted to the {\it Hermitian} parametric chiral random matrices.
In terms of chiral Lagrangian, the difference
is solely in the sign of $\mu^2$, and  the integral formulas  involving 
(original or modified) Bessel functions (\ref{chGOEchGUE}), (\ref{chGSEchGUE}) originated from
the $\sigma$ model action are shared by the two.
Similar complementary treatments were applied for
three-color QCD at real and imaginary {\em isospin} chemical potential,
which correspond to non-Hermitian chiral random matrices \cite{dam1}
and Hermitian chiral random matrices in chGUE-chGUE crossover \cite{dam2,bw,ai}, respectively, 
resulting in a successful determination of $F$.
Combined with the results reported in this paper filling the missing pieces, 
the fact that the Dirac spectral statistics in all three cases 
(SU(2) fundamental+$\mu$, SU(2) adjoint+$\mu$, SU(3) fundamental+$\mu_{\rm iso}$)
agree perfectly with
the predictions from corresponding identical zero-mode-approximated chiral Lagrangians
in both regions of signs of $\mu^2$
constitutes a solid evidence 
for the validity of analytic continuation in the $\mu$ plane,
which is much required for the actual physics, i.e., three-color QCD at real baryon number chemical potential.

We consider the use of imaginary chemical potential has a practical advantage for the following reason.
In order to fit non-Hermitian Dirac spectra to the non-Hermitian random matrix result,
one usually projects the complex eigenvalues either to the real or imaginary axis,
and this projection could blur the fitting. 
Freer two-dimensional motion of complex eigenvalues is likely to lead
to large statistical fluctuation.
On the other hand, confining the eigenvalues within the real axis
averts such issues, yielding a precise fitting shown in Figs.6$\sim$11.

Sharpness of the fitting functions chosen, $p_1(s)$ and $P(s)$, is also in
our advantage.
Note that the use of Wigner surmise (random $2\times2$ matrices) of $p_1(s)$ and $P(s)$
for the parametric random matrices \cite{abps,sbw} 
is an uncontrolled approximation and is not suited for the precise determination of $F$,
especially from fitting in the range of $s\gtrsim2$ \cite{nis2012}.\footnote{%
Take for instance the asymptotic behavior of the  level spacing distribution of GOE,
$\log P(s)\sim -\gamma s^2\ (s\gg1)$.
The Wigner-surmised value $\gamma=\pi/4$ is larger than the exact value $(\pi/4)^2$ by 0.17,
which brings a fatal 100\% error at $s\sim2$.}
The advantage of our treatment is
that exponentially fast, uniform convergence is guaranteed for Nystr\"{o}m-type method
at increasing order.

Our next obvious step is
to include weakly coupled QED in the simulation, rather than the phase noise treated
in this paper.
Our preliminary study shows that Dirac spectra of
SU(2)$\times$U(1) quenched  gauge theory are again fitted well to the parametric random matrix
predictions, and we are currently accumulating numerical data on lattices of larger size
than the current paper.
This two-color QCD + QED model could be of interest to the lattice gauge community
in which the three-color QCD + QED simulation in pursuit of precise measurement of
isospin-related observables has attracted attention recently \cite{bdhiuyz},
although we are exploiting the very difference of symmetry of SU(2)$\times$U(1) 
as compared to SU(3)$\times$U(1).
Other possible extensions are the following:\\

\renewcommand{\labelenumi}{(\roman{enumi})}
\begin{enumerate}
\item
In order to reduce the statistical error of fitting SEDs $p_1(s)$
for which the spectral averaging is not applicable,
one could simultaneously use the 
$k^{\rm th}$ smallest eigenvalues distribution $p_{k}(s), k=2,3,\ldots$ [computable by Eq.(\ref{k-th})]
alongside with $p_1(s)$ for fitting $\rho$ and look for the overlapping of their error bars.
\item
Extension of our treatment to 
Dirac spectrum in a nontrivial SU(2) gauge field topology is straightforward:
on the lattice side one should measure the overlap Dirac spectrum \cite{bw}, 
whereas on the random matrix side the indices of the Bessel $J$ functions
in Eqs.(\ref{chGOEchGUE}) and (\ref{chGSEchGUE}) are to be incremented
by the topological charge $\nu$.
\item
Introduction of dynamical quarks is interesting beyond the obvious reason of
approaching more ``realistic'' models of QCD;
the weakly symmetry-violating U(1) gauge randomness couples to
the SU(2) gauge randomness through the quark loops (fermion determinant) only in that case.
This correlation between perturbed and perturbing can possibly bring nontrivial distortion 
to the relationship between the bare U(1) coupling constant and the
crossover parameter, i.e.~the pion decay constant.
Introduction of finite quark masses on the random matrix side could become cumbersome, but in the light of
individual eigenvalue distributions for 
chGUE-chGUE crossover (the latter of Ref.~\cite{dam2}), the computation is still feasible.
\item 
Our result suggests a possibility of an exotic continuum limit,
in which the rate of decoupling of U(1) gauge field 
is adjusted to the rate of approaching the thermodynamic limit,
while keeping the $\mu^2 F^2$ fixed.
Albeit a rather artificial limit, this procedure might define a theory
in which a nontrivial effect is induced to the interaction of Nambu-Goldstone bosons
by the decoupling gauge interaction through the violation of antiunitary symmetry.
The asymptotic slavery of U(1) is not 
an essential obstacle to this possibility,
because any gauge field in the complex representation, 
namely the fundamental of SU($N'\geq 3$), 
is equally suited as a pertinent perturbation violating the antiunitary symmetry of SU(2).
\end{enumerate}

\noindent
In forthcoming papers we wish to complete the project started here
by covering up the directions listed above.

\begin{acknowledgments}
I thank Taro Nagao for valuable communications at various stages of this work,
and Atsushi Nakamura for helpful discussions and for providing me
with the SU(2) version of his {\sc LatticeToolKit f90} package \cite{ltk}.
I also thank the anonymous referee for accurate suggestions and constructive criticisms 
that helped to rectify and improve the original manuscript.
Discussions during the workshop on ``Field Theory and String Theory'' (YITP-W-12-05) 
at Yukawa Institute for Theoretical Physics at Kyoto University
were useful in the final stage of this work.

All numerical data tables of the level spacing distributions and 
correlation functions,
$p_1^{(\rho)}(s)$, $P^{(\rho)}(s)$,
etc.~plotted in Figs.1$\sim$4
will be provided to interested readers upon email request to the author.
\end{acknowledgments}

\end{document}